\documentclass[namedreferences]{solarphysics}
\usepackage[hyperref,optionalrh]{spr-sola-addons} 
\usepackage{epsfig}          
\usepackage{graphicx}        
\usepackage{color}           
\usepackage{breakurl}        

\newcommand\araa{{ARA\&A}}
\newcommand\apj{{ApJ}}
\newcommand\apjl{{ApJL}}     
\newcommand\apjs{{ApJS}}
\newcommand\aap{{A\&A}}
\newcommand\mnras{{MNRAS}}
\newcommand\solphys{{SoPh}}
\newcommand\ssr{{SSRv}}
\newcommand\grl{{Geophys.~Res.~Lett.}}
\newcommand\jgr{{J.~Geophys.~Res.}}

\chardef\us=`\_

\begin{document}
\begin{article}
\begin{opening}

\title{Solar Cycle Dependence of ICME Composition}

\author[addressref={aff1,aff2},corref,email={hqsong@sdu.edu.cn}]{\inits{H.}\fnm{Hongqiang}~\lnm{Song}}
\author[addressref=aff2]{\inits{L.}\fnm{Leping}~\lnm{Li}}
\author[addressref=aff1]{\inits{Y.}\fnm{Yanyan}~\lnm{Sun}}
\author[addressref=aff1]{\inits{Q.}\fnm{Qi}~\lnm{Lv}}
\author[addressref=aff1]{\inits{R.}\fnm{Ruisheng}~\lnm{Zheng}}
\author[addressref=aff1]{\inits{Y.}\fnm{Yao}~\lnm{Chen}}
\address[id=aff1]{Shandong Key Laboratory of Optical Astronomy and Solar-Terrestrial Environment, Institute of Space Sciences, Shandong University, Weihai 264209, Shandong, China}
\address[id=aff2]{CAS Key Laboratory of Solar Activity, National Astronomical Observatories, Chinese Academy of Sciences, Beijing, 100101, China}
\runningauthor{H. Song et al.}
\runningtitle{Solar Cycle Dependence}

\begin{abstract}
Coronal mass ejections (CMEs) are one of the most energetic explosions in the solar atmosphere, and their occurrence rates exhibit obvious solar cycle dependence with more events taking place around solar maximum. Composition of interplanetary CMEs (ICMEs), referring to the charge states and elemental abundances of ions, opens an important avenue to investigate CMEs. In this paper, we conduct a statistical study on the charge states of five elements (Mg, Fe, Si, C, and O) and the relative abundances of six elements (Mg/O, Fe/O, Si/O, C/O, Ne/O, and He/O) within ICMEs from 1998 to 2011, and find that all the ICME compositions possess the solar cycle dependence. All of the ionic charge states and most of the relative elemental abundances are positively correlated with sunspot numbers (SSNs), and only the C/O ratios are inversely correlated with the SSNs. The compositions (except the C/O) increase with the SSNs during the ascending phase (1998--2000 and 2009--2011) and remain elevated during solar maximum and descending phase (2000--2005) compared to solar minimum (2007--2009). The charge states of low-FIP (first ionization potential) elements (Mg, Fe, and Si) and their relative abundances are correlated well, while no clear correlation is observed between the C$^{6+}$/C$^{5+}$ or C$^{6+}$/C$^{4+}$ and C/O. Most interestingly, we find that the Ne/O ratios of ICMEs and slow solar wind have the opposite solar cycle dependence.
\end{abstract}


\end{opening}

\section{Introduction}

Coronal mass ejections (CMEs) are one of the energetic explosions occurred in the solar atmosphere \citep{forbes00,chenpengfei11,webb12}, which are called interplanetary CMEs (ICMEs) after they leave the corona. ICMEs can propagate at high speed in the interplanetary space \citep{liuying14,manchester17} and induce strong geomagnetic storms when they arrive in the Earth's magnetosphere \citep{gosling91,zhangjie07,shenchenglong17}. CMEs result from eruptions of magnetic flux ropes that can form prior to \citep{patsourakos13,chengxin14a} and during \citep{song14a,ouyang15} solar eruptions. Studies demonstrate that both magnetic reconnection \citep{linjun00,zhangjie01,qiujiong04,maricic07,miklenic09,zhuchunming20} and ideal magnetohydrodynamic instability of flux ropes \citep{forbes95,chenyao07a,song13,song15a,song18a,chengxin20} can contribute to the CME acceleration.

The counterpart of CME flux ropes in the interplanetary space is called magnetic cloud \citep{burlaga81} that is only detected in about one-third ICMEs near 1 au \citep[e.g.,][]{chiyutian16}. Researchers have suggested that the geometric selection effect decides whether the magnetic-cloud features are detected or not \citep{zhangjie13,xie13}, which is further supported by recent studies \citep{maricic20,song20a}. The in situ measurements of ICMEs can provide information on their velocity, density, temperature, magnetic field, as well as composition, including the charge states and the elemental abundances of different ions, which are very helpful to investigate CMEs \citep{song20b}.

The elemental abundances in ICMEs do not change during their transit to 1 au, while the ionic fractions for a given chemical element may vary substantially near the Sun. For simplicity, let us assume that the velocity difference between species in different charge states is not that important for determining the ionic fractions \citep[see e.g.,][]{chenyao03}. This then makes it meaningful to follow a fluid parcel moving with the bulk velocity $\vec{v}$ and varying at a dynamic timescale $\tau_{\rm dyn} = 1/|D/Dt|$, where $D/Dt = \partial/\partial t + \vec{v}\cdot\nabla$ is the material derivative. The so-called ionization equilibrium is realized when $\tau_{\rm dyn}$ is much longer than the timescales charactering the ionizaiton/recombination processes ($\tau_{\rm ion}$). If accepting that the ionization is primarily due to electron impact (and possibly autoionizations) and the recombination largely involves radiative and dielectronic processes, then the ionization/recombination frequencies ($\nu_{\rm ion} \equiv 1/\tau_{\rm ion}$) are proportional to the electron density $N_e$ as well as the ionization and recombination rate coefficients, which in turn depend essentially only on the electron temperature ($T_e$)\citep[e.g.,][]{owocki83,DelZanna18,shimijie19,libo20}. Ionization equilibrium is expected when a fluid parcel just leaves its source regions given the pertinent high values of $N_e$. However, when a fluid parcel moves through the first several solar radii, $N_e$ decreases rapidly with time (or equivalently with heliocentric distance), making $\tau_{\rm dyn}$ (much) shorter than $\tau_{\rm ion}$ beyond some distance. \cite{rivera19a} have visualized the departures from ionization equilibrium for a number of ions within ICMEs near the Sun. The end result is that the ionization fractions do not vary any longer, \textit{i.e.}, freeze-in. This means that the ionic fractions measured in situ are largely determined by the physical parameters near the Sun where freeze-in occurs \citep{rakowski07,lynch11,gruesbeck11,gruesbeck12,song15b,song15c}, which have been further used to infer the eruption process of flux ropes \citep{song16,wangwensi17,huangjia18}. It should be noted that the different structures of CMEs have different freeze-in history \citep{gruesbeck12,rivera19b}, which leads to the complex nature of ICMEs and makes the interpretation of ions in situ less straightforward.

The O$^{7+}$/O$^{6+}$ and C$^{6+}$/C$^{5+}$ have long been adopted to differentiate the solar wind sources \citep{zurbuchen02,zhaoliang17}, and \cite{landi12a} showed that the C$^{6+}$/C$^{4+}$ is more sensitive to indicate the electron temperatures. The charge states of C$^{4\sim6+}$ and O$^{5\sim7+}$ freeze-in close to the Sun, and the formation temperature of C$^{4+}$, C$^{6+}$, O$^{6+}$, and O$^{7+}$ covers a large range \citep{landi12a,landi12b}. Therefore, their freeze-in abundances do not change drastically prior to freeze-in and remain good representations of the source region of solar wind. Contrary to the charge state ratios of O and C, the average charge state of Fe ($<$Q$_{Fe}$$>$) acts as the sensitive indicator of electron temperatures in the corona \citep[e.g.,][]{lepri13}. Fe charge states freeze-in at a larger distance range from 1.2 $R_\odot$ to beyond 5 $R_\odot$ in the solar wind and farther for ICMEs. Thus the Fe charge states become less representative of source region, while are more appropriate for analyzing CME eruption process \citep{song16,wangwensi17}.

Observations have found that the abundances of elements with a low first ionization potential (FIP) are often enhanced over the high-FIP elements in the solar upper atmosphere and solar wind. The separation between low and high FIP is $\sim$10 eV, and Mg, Fe and Si are considered as low-FIP elements while O, Ne and He are considered as high-FIP elements. \cite{young05} reported that the Mg/Ne abundance ratio in the transition region of the Sun is enhanced over the photospheric value by a factor less than 2, and \cite{feldman98} found that the low-FIP elements are enriched by about a factor of 4 in the corona above the quiet equatorial region. This abundance anomaly pattern is referred to as the``FIP effect" \citep{laming15}. The solar wind plasma is accelerated in the solar corona and its elemental abundances are expected to be consistent with that of the corona, rather than the photosphere \citep[e.g.,][]{schmelz12,zhaoliang17}. The FIP effect can be employed to diagnose the origin of ICME plasmas \citep{zurbuchen16,song17a,fuhui20}. Therefore, several studies have been conducted on the characteristics and applications of ICME composition \citep[and references therein]{song20b}. For example, \cite{song17a} analyzed the elemental abundances of an erupted filament within an ICME near 1 au and found that its relative abundances are close to the corresponding photospheric values. This does not support that the filament plasma originates from the corona.

\cite{owens18} analyzed the charge states of C, O, and Fe within 215 ICMEs, including 97 magnetic clouds and 118 non-cloud events. The work reported that magnetic clouds exhibit higher ionic charge states than non-magnetic clouds. In addition, statistical results demonstrated that fast magnetic clouds have higher charge states and relative elemental abundances (except the C/O) than slow ones \citep{owens18,huangjin20}, and cold prominence plasmas with lower charge states can be detected within ICMEs near 1 au \citep{lepri10,gilbert12,sharma12,wangjiemin18,fengxuedong18}. \cite{zurbuchen16} performed a comprehensive analysis of the elemental abundances of 310 ICMEs on Richardson \& Cane's catalog from 1998 March to 2011 August \citep{richardson10}. They reported that the abundances of low-FIP elements within ICMEs exhibit a systematic increase compared to the solar wind, and the ICMEs with elevated iron charge states possess higher FIP fractionation than the other ICMEs.

Solar cycle has an average period of $\sim$11 years, which is well represented by the sunspot numbers (SSNs). Remote observations report that the Sun's evolution and variation over the cycle are very obvious. For instance, the structure of solar corona varies with solar cycle from a near spherical symmetry at solar maximum to an axial dipole at minimum \citep{dikpati16}. The numbers of active regions in the solar atmosphere decrease gradually from solar maximum to minimum, which might reflect the underlying physical driver of the variations -- the solar dynamo \citep{charbonneau20}. More solar activities, such as CMEs, flares and filament eruptions, occur in active regions around solar maximum \citep[e.g.,][]{chenpengfei11}. As both slow solar wind and ICMEs originate from the quiet region and active region, which have different physical properties, such as temperature, density, as well as magnetic field structure, and the detected slow wind and ICMEs near 1 au are more from the active region around solar maximum compared to minimum, the measured composition of solar wind and ICMEs should exhibit the solar cycle dependence.

Previous studies showed that both the charge states and elemental abundances of solar wind (with the exclusion of ICME periods) have a solar cycle dependence \citep{lepri13,shearer14,zhaoliang14,zhaoliang17}. A recent study revealed that the Mg/O, He/O, and the average charge states of C and O within ICMEs exhibit the solar cycle dependence \citep{guchaoran20} similar to the solar wind. As ICMEs and solar wind result from different mechanisms \citep{chenpengfei11,abbo16}, a more comprehensive statistical study on the solar cycle variation of ICME composition is necessary to examine whether some differences exist between ICMEs and solar wind and to infer more knowledge about CMEs and/or FIP effect. This is the major motivation of this paper. We introduce the data and ICME catalog in Section 2, present the statistical results in Section 3 and give the discussion in Section 4. Section 5 is the summary.

\section{Data and ICME catalogs}
The compositional data in this study are obtained from the Solar Wind Ion Composition Spectrometer \citep[SWICS;][]{gloeckler98} aboard the \textit{Advanced Composition Explorer (ACE)}, which was launched in 1997 and orbits around the L1 point. SWICS is optimized for low-noise measurements of the solar wind composition. It identifies a solar wind ion through a combination of three independent measurements, \textit{i.e.}, an electrostatic selection of its energy-per-charge (E/q), a time-of-flight measurement of its speed, and a measurement of its total energy \citep{gilbert12}. SWICS can provide the charge-state distributions and abundances of $\sim$10 elements, and the newly released SWICS 1.1 level 2 data \citep{shearer14} are used in our analysis, which can be downloaded from the \textit{ACE} science center\footnote{http://www.srl.caltech.edu/ACE/ASC/level2/index.html}. The yearly average SSNs are from the Solar Influence Data Center of the Royal Observatory of Belgium\footnote{http://www.sidc.be/silso/home}.

Experts in ICME fields have provided several complete and reliable catalogs based on measurements of the \textit{ACE} \citep[RC catalog]{richardson10}, \textit{WIND} \citep{chiyutian16,nieves18}, and the \textit{Solar Terrestrial Relations Observatory} \citep{jianlan18}. Every catalog contains the ejecta boundaries of each ICME, which can be adopted to analyze the composition of ICME ejecta. As the ICME compositional data analyzed in this paper are provided by the SWICS aboard the \textit{ACE}, we choose to use the RC catalog\footnote{http://www.srl.caltech.edu/ACE/ASC/DATA/level3/icmetable2.html}.

The optimal SWICS data are available online from 1998 February to 2011 August, during which 319 ICMEs in total appear in the RC catalog. The histograms in Figure 1 display the yearly numbers of ICMEs from 1996 to 2011 with the gray denoting the events occurred between 1998 February and 2011 August. Note that there are 36 and 32 ICMEs in 1998 and 2011, respectively, while no optimal SWICS data for 5 events in 1998 and 20 events in 2011. The data gap only appears in a few events, which are not included automatically when calculating the average values.
\begin{table*}
\caption{The analyzed ICME number, average value (mean) and standard deviation (stddev) of every ICME compositional parameter in each year from 1998 to 2011.}
\tabcolsep=2pt
\begin{tabular}{cccccccccccccccc}
\hline
\hline
 & Year    & 1998 & 1999 & 2000 & 2001 & 2002 & 2003 & 2004 & 2005 & 2006 & 2007 & 2008 & 2009 & 2010 & 2011 \\
 & Number  & 31   & 33   & 51   & 48   & 26   & 22   & 21   & 31   & 13   & 2    & 3    & 11   & 15   & 12   \\
\hline
$<$Q$_{Mg}$$>$    &mean  & 9.271 & 9.255 & 9.392 & 9.347 & 9.363 & 9.374 & 9.365 & 9.274 & 9.236 & 8.975 & 8.627 & 8.640 & 9.014 & 9.125\\
     &stddev& 0.237 & 0.200 & 0.222 & 0.288 & 0.214 & 0.245 & 0.276 & 0.285 & 0.242 & 0.082 & 0.369 & 0.190 & 0.183 & 0.353\\
$<$Q$_{Fe}$$>$    &mean  & 11.25 & 11.26 & 11.87 & 11.56 & 11.74 & 12.32 & 12.26 & 11.82 & 11.33 & 10.38 & 9.798 & 9.473 & 10.41 & 11.07\\
     &stddev& 1.326 & 1.118 & 1.507 & 1.602 & 1.377 & 1.477 & 1.547 & 1.458 & 1.494 & 0.114 & 0.493 & 0.482 & 0.919 & 1.515\\
$<$Q$_{Si}$$>$    &mean  & 9.816 & 9.867 & 10.09 & 9.984 & 10.15 & 10.22 & 10.17 & 9.873 & 9.737 & 9.263 & 8.807 & 8.572 & 9.275 & 9.598\\
     &stddev& 0.615 & 0.485 & 0.571 & 0.721 & 0.583 & 0.639 & 0.713 & 0.639 & 0.589 & 0.305 & 0.547 & 0.283 & 0.465 & 0.664\\
C$^{6+}$/C$^{5+}$&mean  & 1.178 & 1.565 & 1.335 & 1.659 & 2.171 & 1.409 & 1.995 & 1.297 & 1.632 & 0.690 & 0.953 & 0.766 & 1.024 & 1.547\\
     &stddev& 0.633 & 0.865 & 0.928 & 1.148 & 1.132 & 0.994 & 1.699 & 0.808 & 0.758 & 0.225 & 0.439 & 0.354 & 0.509 & 1.255\\
C$^{6+}$/C$^{4+}$&mean  & 5.170 & 7.772 & 6.522 & 7.353 & 9.554 & 6.414 & 6.902 & 4.757 & 6.635 & 2.256 & 2.215 & 2.126 & 3.517 & 7.480\\
     &stddev& 3.180 & 6.173 & 4.896 & 5.753 & 6.441 & 4.316 & 5.952 & 3.422 & 4.173 & 0.459 & 1.098 & 1.193 & 2.118 & 7.649\\
O$^{7+}$/O$^{6+}$&mean  & 0.563 & 0.578 & 0.697 & 0.712 & 0.772 & 0.790 & 0.733 & 0.571 & 0.486 & 0.207 & 0.176 & 0.114 & 0.308 & 0.440\\
     &stddev& 0.318 & 0.262 & 0.420 & 0.494 & 0.351 & 0.793 & 0.469 & 0.490 & 0.459 & 0.041 & 0.050 & 0.036 & 0.162 & 0.264\\
Mg/O &mean  & 0.227 & 0.228 & 0.257 & 0.259 & 0.244 & 0.252 & 0.253 & 0.226 & 0.174 & 0.154 & 0.146 & 0.142 & 0.169 & 0.204\\
     &stddev& 0.074 & 0.043 & 0.092 & 0.086 & 0.073 & 0.177 & 0.088 & 0.103 & 0.063 & 0.018 & 0.022 & 0.020 & 0.042 & 0.043\\
Fe/O &mean  & 0.224 & 0.213 & 0.282 & 0.273 & 0.263 & 0.297 & 0.237 & 0.220 & 0.150 & 0.135 & 0.159 & 0.177 & 0.184 & 0.223\\
     &stddev& 0.105 & 0.091 & 0.137 & 0.103 & 0.109 & 0.235 & 0.103 & 0.096 & 0.048 & 0.016 & 0.064 & 0.095 & 0.062 & 0.143\\
Si/O &mean  & 0.205 & 0.205 & 0.239 & 0.242 & 0.223 & 0.231 & 0.216 & 0.211 & 0.164 & 0.163 & 0.131 & 0.143 & 0.170 & 0.199\\
     &stddev& 0.064 & 0.039 & 0.064 & 0.060 & 0.050 & 0.145 & 0.054 & 0.067 & 0.037 & 0.008 & 0.029 & 0.029 & 0.035 & 0.049\\
C/O  &mean  & 0.514 & 0.579 & 0.470 & 0.521 & 0.541 & 0.493 & 0.556 & 0.529 & 0.588 & 0.491 & 0.599 & 0.598 & 0.534 & 0.540\\
     &stddev& 0.155 & 0.187 & 0.117 & 0.124 & 0.138 & 0.167 & 0.164 & 0.132 & 0.151 & 0.125 & 0.084 & 0.072 & 0.118 & 0.152\\
Ne/O &mean  & 0.188 & 0.179 & 0.215 & 0.210 & 0.211 & 0.249 & 0.227 & 0.199 & 0.181 & 0.151 & 0.140 & 0.158 & 0.159 & 0.178\\
     &stddev& 0.065 & 0.052 & 0.106 & 0.106 & 0.070 & 0.171 & 0.100 & 0.097 & 0.080 & 0.018 & 0.008 & 0.026 & 0.027 & 0.052\\
He/O &mean  & 104.2 & 117.5 & 104.9 & 113.5 & 132.6 & 110.6 & 118.0 & 97.95 & 89.03 & 53.89 & 78.12 & 65.40 & 73.04 & 109.3\\
     &stddev& 48.87 & 50.29 & 41.48 & 60.19 & 58.16 & 75.57 & 49.47 & 48.86 & 42.44 & 32.72 & 28.82 & 15.78 & 25.06 & 82.06\\
\hline
\hline
\end{tabular}
\end{table*}

To demonstrate the solar cycle dependence of ICME numbers, the yearly average SSNs are plotted together in Figure 1 as shown with the black line. To evaluate the correlation between the yearly numbers of ICMEs and sunspots quantitatively, we calculate their linear Pearson correlation coefficient with the IDL routine $correlate.pro$. This coefficient is defined in statistics as the measurement of the strength of the relationship between two variables and their association with each other. It has a value between -1 and 1, and -1 (1) indicates a perfectly negative (positive) linear correlation between the two variables, while 0 indicates no linear correlation between them. Usually, the absolute values 0.8$\sim$1.0 (0.6$\sim$0.8) represent a fairly strong (strong) relationship, and 0.4$\sim$0.6 (0.2$\sim$0.4), a moderate (weak) relationship. The correlation coefficient between the numbers of ICMEs and sunspots is 0.85, illustrating fairly strong correlation between them.

\section{Statistical results}
To analyze the solar cycle dependence of ICME composition, we first calculate the average value of every composition within each ICME with the two-hour time resolution SWICS data \citep{shearer14}, and then get the yearly means and standard deviations of every composition based on all ICMEs in each year. All of the calculated results are listed in Table 1, which also displays the analyzed ICME number in each year in the second row.

\subsection{The ionic charge states}
The yearly means of $<$Q$_{Mg}$$>$, $<$Q$_{Fe}$$>$, $<$Q$_{Si}$$>$, C$^{6+}$/C$^{5+}$, C$^{6+}$/C$^{4+}$, O$^{7+}$/O$^{6+}$, and SSNs are displayed sequentially in Figures 2(a)--(g). The red vertical lines in both Figures 2 and 3 represent the standard deviations. Figure 2 illustrates that the ionic charge states increase with SSNs during the ascending phase (1998-2000 and 2009--2011) and remain elevated during solar maximum and descending phase (2000--2005) compared to the minimum phase (2007-2009). The yearly means of $<$Q$_{Mg}$$>$ ($<$Q$_{Fe}$$>$, $<$Q$_{Si}$$>$, C$^{6+}$/C$^{5+}$, C$^{6+}$/C$^{4+}$, and O$^{7+}$/O$^{6+}$) drop up to $\sim$0.77 (2.85, 1.65, 1.48, 7.43, and 0.68) from solar maximum to minimum. The correlation coefficients between the yearly means of charge states and SSNs are 0.74, 0.64, 0.74, 0.63, 0.80, and 0.80 sequentially, which are also displayed in Panels (a)--(f) and demonstrate quantitatively that all of the charge states possess the solar cycle dependence.

The similar solar cycle dependence of ionic charge states also exists in the slow solar wind. The left and right blue numbers in Figures 2(b) and (d)-(f) denote the corresponding means of slow wind parameters during solar maximum (2000-2002) and minimum (2008--2009) \citep{lepri13}, respectively. As CMEs originate from active regions and quiet regions, which are also the sources of slow wind, it is reasonable to see the similar solar cycle variation. However, Figure 2 shows that the average charge states of ICMEs are obviously higher than those of slow wind around solar maximum, while their differences are relatively slight around minimum. In addition, the correlation coefficient between the C$^{6+}$/C$^{4+}$ and SSNs is 0.80, which is larger than the coefficient between C$^{6+}$/C$^{5+}$ and SSNs, supporting that C$^{6+}$/C$^{4+}$ is a more sensitive indicator of coronal electron temperature \citep{landi12a}.

\subsection{The elemental abundances}
Figure 3 shows the yearly means for Mg/O, Fe/O, Si/O, C/O, Ne/O, He/O, and SSNs from top to bottom panels sequentially. The left and right blue numbers in Panels (b)--(f) also denote the corresponding means of slow wind parameters around solar maximum and minimum, respectively \citep{lepri13,shearer14}. We can see that most relative abundances (except the C/O) increase with the SSNs during the ascending phase (1998-2000 and 2009--2011) and remain relatively elevated during solar maximum and descending phase (2000--2005) compared to minimum phase (2007-2009). The yearly means of Mg/O (Fe/O, Si/O, Ne/O, and He/O) decrease by a factor of up to $\sim$45\% (55\%, 46\%, 44\%, and 59\%) from solar maximum to minimum. The correlation coefficients between these elemental abundances and SSNs are 0.85, 0.83, 0.87, 0.65, and 0.80, respectively, demonstrating the strong solar cycle dependence.

Conversely, the C/O ratios show negative correlation with SSNs as presented in Panel (d), and the yearly means of C/O increase by a factor of up to $\sim$27\% from solar maximum to minimum. This agrees with the variation trend in the slow wind \citep[e.g.,][]{lepri13}. Most interestingly, the Ne/O ratios of ICMEs and slow wind possess the opposite solar cycle dependence, i.e., the Ne/O values of slow wind increase from solar maximum to minimum \citep{shearer14,zhaoliang17}. These will be discussed later in Section 4.

As mentioned, the relative elemental abundances in the corona are different from their photospheric values, which can be described with FIP bias factor. The factor refers to the ratio of relative abundance (e.g., Fe/O) in the corona over the abundance in the photosphere. Researchers have provided several data sets on the photospheric abundances \citep[e.g.,][]{grevesse98,asplund09,caffau11}, and some differences exist in different literatures. We adopt the photospheric abundances of \cite{grevesse98}, as it has been used to study the FIP bias of ICMEs \citep{zurbuchen16} and solar wind \citep{zhaoliang17}, which is convenient for comparison between our study and their results. The curves in Panels (a)--(f) also describe the variations of FIP bias factors (see the right ordinates). The relative abundances in the photosphere \citep{grevesse98} and the FIP bias range of ICMEs are also presented with black text in these panels.

Figure 3 shows that the FIP bias factors of low-FIP elements (Mg, Fe, and Si) within ICMEs maintain over 1, and most of the factors of C and He keep beyond and below 1 during the whole solar cycle, respectively, which are consistent with the slow wind \citep{zhaoliang17}. The factors of Ne for ICMEs are beyond (below) 1 around solar maximum (minimum), different from the slow wind which maintains values below 1 \citep{zhaoliang17}. The horizontal blue dotted lines in Panels (d)--(f) denote the corresponding relative abundances in the photosphere (FIP bias factor $=$ 1). \cite{zurbuchen16} reported that the factors of Ne for the ICMEs with (without) high $<$Q$_{Fe}$$>$ are beyond (below) 1, where an ICME is defined as a high $<$Q$_{Fe}$$>$ one if it contained a minimum of 6 hr of plasma with $<$Q$_{Fe}$$>$ larger than 12 \citep{lepri01}. This agrees with our present study as more high $<$Q$_{Fe}$$>$ ICMEs appear around solar maximum \citep{song16}, with their FIP bias factors of Ne beyond 1 \citep{zurbuchen16}.

The statistical results based on fewer events should not be credible, so we do not further divide the yearly ICMEs into fast and slow ones, or magnetic clouds and non-cloud events \citep{owens18}. There were only a few events in 2007 and 2008, while 29 events in total were detected during the solar minimum (2006--2009). The yearly average values of compositional parameters are close in these years, while different from those around solar maximum. Therefore, the overall variation trends of compositions from solar maximum to minimum should not result from the event selection effect. In addition, we conduct a t-test to compare the means of each parameter around solar maximum (2000--2003) and minimum (2006--2009) as some standard deviations are relatively large. The t-test is a type of inferential statistic used to determine if there is a significant difference between the means of two groups. As we want to check if the difference of ICME compositional means between solar maximum and minimum is significant, we conduct a t-test on each parameter (e.g.,$<$Q$_{Fe}$$>$, Fe/O, etc.) between two groups (solar maximum and minimum) through the IDL routine $tm\_test.pro$. The function computes the Student’s T-statistic and the probability (p-value) that two sample populations X and Y have significantly different means. The p-value is in the interval [0.0, 1.0], and a small value (less than 0.05 or 0.01) indicates that X and Y have significantly different means. Our t-test results show that all the p-values are less than 0.01, which illustrates the means of every composition around solar maximum and minimum have significant difference at the 99\% confidence level.

\subsection{Correlations between the relative abundances and charge states of heavy ions}
Figure 4 presents the scatter plots of the yearly means of relative abundances and charge states of heavy ions within ICMEs. As both the abundances and charge states of low-FIP elements are correlated well with the SSNs, it is natural to expect that correlation exists between the Mg/O (Fe/O and Si/O) and the $<$Q$_{Mg}$$>$ ($<$Q$_{Fe}$$>$, and $<$Q$_{Si}$$>$) as displayed in Panels (a)--(c). The correlation coefficients are 0.91, 0.76, and 0.91 for Mg, Fe, and Si, respectively. However, no clear correlation is observed between the C/O and the C$^{6+}$/C$^{5+}$ or C$^{6+}$/C$^{4+}$ as shown in Panels (d) and (e).

\section{Discussion}
The ionic charge states within ICMEs are mainly governed by the thermodynamic evolution of the electrons in CMEs up to freeze-in distance of each ion \citep{owocki83,landi12a,landi12b,song16}. More CMEs are associated with energetic flares around solar maximum \citep[e.g.,][]{kouyuankun20}, indicating higher electron temperatures and then more elevated ionic charge states due to magnetic reconnections \citep[e.g.,][]{lepri04,song16}. The elevated charge states generated in the current sheets connecting magnetic flux ropes and flare loops can flow into CMEs during eruptions \citep{song16,wangwensi17}. Therefore, it is reasonable that the charge states within ICMEs are positively correlated with the SSNs, which agree with the expectations of current CME models \citep{mikic94,antiochos99,linjun00,moore01,fanyuhong07,chenpengfei08}. However, it is not so straightforward to understand the variations of relative elemental abundances as they can be positively or inversely correlated with the SSNs and some differences exist between ICMEs and solar wind.

According to a model of the FIP effect \citep{laming04,laming12}, the elemental fractionation results from the ponderomotive force acting on the ions from Alfv\'{e}n waves in the chromosphere. More Alfv\'{e}n waves could be generated in the corona through magnetic reconnections around solar maximum, and this might be employed to explain the solar cycle dependence of relative abundances of low-FIP elements in both ICMEs and solar wind \citep{lepri13,zhaoliang17}. This can also explain why the coronal temperatures are higher around solar maximum compared to minimum as both magnetic reconnections and Alfv\'{e}n waves can contribute to the coronal heating \citep{cranmer19}. Therefore, the relative elemental abundances and ionic charge states (except the C/O and C$^{6+}$/C$^{5+}$ or C$^{6+}$/C$^{4+}$) are correlated well as both of them correlate with the SSNs well though they are determined by different physical processes.

\subsection{The negative correlation between C/O and SSNs}

The high-FIP elements exhibit some different behaviors. As the FIP of C (11.26 eV) is lower but close to that of O (13.62 eV), it is expected that the overall C/O ratios show positive correlations with the SSNs, instead of negative correlation in both ICMEs and slow wind as shown in Figure 3(d). \cite{zhaoliang17a} reported an anomalous composition in the slow wind, which has the same charge state composition compared to the normal slow wind but the abundances of fully charged species decrease, leading to the depletion of He and C elemental abundances. The anomalous phenomenon appears more frequent around solar maximum. \cite{kocher17} conducted a statistical study on the charge state composition of each ICME from 1998 to 2011. They found that 44\% of the ICMEs possess the similar anomalous composition and labeled them ``depleted ICMEs". The number of depleted ICMEs is most abundant around solar maximum and almost no ones are found around minimum. They also reported that the C/O ratios are lower in nearly all the depleted ICMEs compared to the surrounding plasma. The anomalous phenomenon in both solar wind \citep{zhaoliang17a} and ICMEs \citep{kocher17} seems to be one factor that changes the correlation between the C/O ratios and SSNs.

To examine the validity of the above explanation, we select 111 non-depleted ICMEs from the 319 ICMEs according to a relatively strict criterion, which requires no any data points within the ICMEs exhibiting the anomalous composition, i.e., $log_{10}R<0.15$, where $R=(C^{6+}/C^{5+})/(O^{7+}/O^{6+})$ \citep{kocher17}. The yearly means of C/O ratios within non-depleted ICMEs and SSNs exhibit a positive correlation (not shown) with the correlation coefficient being 0.40. This is contrary to the negative correlation when all the ICMEs are involved as we have known, and supports that the anomalous phenomenon in ICMEs can influence the solar cycle dependence of C/O ratios. The He abundances are significantly higher than the other heavy elements, and the anomalous phenomenon might not be able to change their overall variation trend throughout the solar cycle.

\subsection{The different solar cycle dependence of Ne/O within ICMEs and solar wind}

The most interesting phenomenon is that the Ne/O ratios show positive and negative correlations with SSNs in ICMEs and slow wind \citep{shearer14,zhaoliang17}, respectively. This is different from the other elements that keep consistent cycle dependence in ICMEs and slow wind no matter positive or negative correlations. The Ne has higher FIP (21.56 eV) than O, so it is expected that the Ne/O ratios are lower around solar maximum according to the above FIP effect model \citep{laming04,laming12}, which should be the situation in the solar wind.

\cite{zurbuchen16} found that the Ne/O correlates with the Fe/O well in the ICMEs with elevated Fe charge states, and we find that the ionic charge states within ICMEs are higher around solar maximum as displayed in Figure 2. Thus it is reasonable that the Ne/O ratios within ICMEs are higher around solar maximum as shown in Figure 3(e). However, the Fe charge states in the slow wind are also higher \citep{lepri13} but their Ne/O ratios are lower \citep{shearer14} around maximum. This implies that ICMEs, especially those with elevated charge states, \textit{i.e.}, associated with strong flares, might suffer different element ionization and fractionation process compared to the solar wind. Here we discuss one potential mechanism that leads to the enhanced Ne/O ratios within ICMEs around solar maximum.

The enhanced Ne/O ratios such as $\sim$0.25 and $\sim$0.32 have been reported in flaring solar plasma \citep{schmelz93,ramaty95}, which are close to the ICME Ne/O values around solar maximum as presented in Figure 3(e) and larger than the abundances in the corona and solar wind \citep{ramaty95,shearer14,zhaoliang17}. The photoionization has been proposed as one possible mechanism to explain the enhanced Ne/O ratios \citep{shemi91,schmelz93}. \cite{shemi91} suggested that the pre-flare soft X-ray can penetrate through the chromosphere and create a slab-like region of non-thermal ionization ratios at the chromosphere base. As the photoionization cross section ratios of O and Ne are 4:9 \citep{yeh85} and the photoionized O can recombine through the efficient charge transfer reactions with the neutral H, the Ne ions are mixed mainly with the neutral O in the slab region. Then the photoionized Ne ions are selected with the thermally ionized low-FIP elements together for preferential transfer to the upper atmosphere.

The detailed calculations supported that the flare ionization can provide sufficient Ne ions in the slab region \citep{shemi91}. \cite{shemi91} concluded that a long-duration ($\ge$30 min) flare, especially flares with a long and intense pre-flare phase, could produce high Ne/O ratios in the flaring region. Given the diffusion and other mechanisms suggested for the quiet-Sun stage could not explain the ion-neutral separation during the mass transport from the slab region to the upper atmosphere in a short-time scale. \cite{shemi91} further proposed a plausible way, where magnetic activity causes the ion-neutral separation either via magnetic field reconnections or by shearing motions. Through this scenario, the plasma with enhanced Ne/O ratios in the flaring region (associated with active-region flares usually around solar maximum) can flow into CMEs \citep{lepri04}, corresponding to the enhanced Ne/O ratios within ICMEs.

\subsection{The large standard deviation}
The elemental compositions within ICMEs are determined by many factors, such as the properties of source regions and the detailed eruption processes \citep{song20b}. In the meantime, many CMEs possess the three-part structure \citep{vourlidas13,song17b,song19a,song19b} that result from filament or hot channel eruptions. The different parts of CMEs could contain different FIP biases and thermodynamic evolution from the Sun. All of these lead to the compositions within ICMEs can vary significantly from case to case. In addition, the measured results are also influenced by the spacecraft trajectory crossing ICMEs as the compositions are not uniform within one ICME. Therefore, it is understandable that the standard deviations of ICME composition are large. Our study demonstrates that there exists the overall solar cycle dependence of ICME composition, which makes sense as the coronal properties evolve from solar maximum to minimum.

\section{Summary}
In this paper, we conducted a statistical study on the solar cycle dependence of the charge states (including $<$Q$_{Mg}$$>$, $<$Q$_{Fe}$$>$, $<$Q$_{Si}$$>$, C$^{6+}$/C$^{5+}$, C$^{6+}$/C$^{4+}$, and O$^{7+}$/O$^{6+}$) and the relative elemental abundances (including Mg/O, Fe/O, Si/O, C/O, Ne/O, and He/O) within ICMEs measured by \textit{ACE} spacecraft from 1998 to 2011. Our major findings can be summarized as follows.

1. All the ICME compositions exhibit the solar cycle dependence. The t-test demonstrates that all the parameters have significantly different means from solar maximum to minimum. All of the ionic charge states and most of the relative elemental abundances (except the C/O) are positively correlated with the SSNs.

2. The compositions (except the C/O) increase with the SSNs during the ascending phase (1998-2000 and 2009--2011) and remain elevated during solar maximum and descending phase (2000--2005) compared to minimum phase (2007--2009).

3. Several of the relative abundances in ICMEs and slow wind exhibit the similar solar cycle dependence, while the Ne/O ratios within ICMEs display the opposite variation trend compared to the slow wind \citep{shearer14,zhaoliang17} from solar maximum to minimum.

4. The ionic charge states and relative abundances within ICMEs are correlated well for low-FIP elements such as Mg, Fe, and Si, while no clear correlation is observed between the C$^{6+}$/C$^{5+}$ or C$^{6+}$/C$^{4+}$ and the C/O.

\begin{acknowledgements}
We thank the referee for his/her constructive comments and suggestions which helped to improve the original manuscript considerably. We acknowledge the use of ICME catalog provided by Richardson \& Cane and the use of data from \textit{ACE} mission. Hongqiang Song thanks Drs. Liang Zhao (University of Michigan), Bo Li (Shandong University), Hui Tian (Peking University), and Liang Guo (a statistician at Shandong University) for their helpful discussions. This work is supported by the NSFC grants U2031109, 11790303 (11790300), and 12073042. Hongqiang Song is also supported by the CAS grants XDA-17040507 and the open research program of the CAS Key Laboratory of Solar Activity KLSA202107.
\end{acknowledgements}

%
%


\begin{thebibliography}{100}
\ifx\bisbn     \undefined \def\bisbn  #1{ISBN #1}\fi
\ifx\binits    \undefined \def\binits#1{#1}\fi
\ifx\bauthor   \undefined \def\bauthor#1{#1}\fi
\ifx\batitle   \undefined \def\batitle#1{#1}\fi
\ifx\bjtitle   \undefined \def\bjtitle#1{\textit{#1}}\fi
\ifx\bvolume   \undefined \def\bvolume#1{\textbf{#1}}\fi
\ifx\byear     \undefined \def\byear#1{#1}\fi
\ifx\bissue    \undefined \def\bissue#1{#1}\fi
\ifx\bfpage    \undefined \def\bfpage#1{#1}\fi
\ifx\blpage    \undefined \def\blpage #1{#1}\fi
\ifx\burl      \undefined \def\burl#1{#1}\fi
\ifx\href      \undefined \def\href#1#2{#2}\fi
\ifx\betal     \undefined \def\betal{et al.}\fi
\ifx\bctitle   \undefined \def\bctitle#1{#1}\fi
\ifx\beditor   \undefined \def\beditor#1{#1}\fi
\ifx\bbtitle   \undefined \def\bbtitle#1{\textit{#1}}\fi
\ifx\bedition  \undefined \def\bedition#1{#1}\fi
\ifx\bseriesno \undefined \def\bseriesno#1{\textbf{#1}}\fi
\ifx\blocation \undefined \def\blocation#1{#1}\fi
\ifx\bsertitle \undefined \def\bsertitle#1{\textit{#1}}\fi
\ifx\bsnm      \undefined \def\bsnm#1{#1}\fi
\ifx\bsuffix   \undefined \def\bsuffix#1{#1}\fi
\ifx\bparticle \undefined \def\bparticle#1{#1}\fi
\ifx\barticle  \undefined \def\barticle#1{}\fi
\ifx\binstitute  \undefined \def\binstitute#1{#1}\fi
\ifx\bpublisher  \undefined \def\bpublisher#1{#1}\fi
\ifx\doiurl    \undefined \def\doiurl#1{\href{#1}{DOI}}\fi
\makeatletter
\def\safeHref#1#2#3{\in@{http}{#2}\ifin@\href{#2}{#3}\else\href{#1#2}{#3}\fi}
\makeatother
\ifx\adsurl    \undefined
  \def\adsurl#1{\safeHref{https://ui.adsabs.harvard.edu/abs/}{#1}{ADS}}\fi
\ifx\arxivurl  \undefined
  \def\arxivurl#1{\safeHref{http://arxiv.org/abs/}{#1}{arXiv}}\fi
\ifx\botherref \undefined \def\botherref#1{}\fi
\ifx\url       \undefined \def\url#1{#1}\fi
\ifx\bchapter  \undefined \def\bchapter#1{}\fi
\ifx\bbook     \undefined \def\bbook#1{}\fi
\ifx\bcomment  \undefined \def\bcomment#1{#1}\fi
\ifx\oauthor   \undefined \def\oauthor#1{#1}\fi
\ifx\citeauthoryear \undefined\def \citeauthoryear#1{#1}\fi
\def\endbibitem {}
\ifx\bconflocation  \undefined \def\bconflocation#1{#1} \fi

\bibitem[\protect\citeauthoryear{{Abbo} et~al.}{2016}]{abbo16}
\begin{barticle}
\bauthor{\bsnm{{Abbo}}, \binits{L.}},
\bauthor{\bsnm{{Ofman}}, \binits{L.}},
\bauthor{\bsnm{{Antiochos}}, \binits{S.K.}},
\bauthor{\bsnm{{Hansteen}}, \binits{V.H.}},
\bauthor{\bsnm{{Harra}}, \binits{L.}},
\bauthor{\bsnm{{Ko}}, \binits{Y.-K.}},
\bauthor{\bsnm{{Lapenta}}, \binits{G.}},
\bauthor{\bsnm{{Li}}, \binits{B.}},
\bauthor{\bsnm{{Riley}}, \binits{P.}},
\bauthor{\bsnm{{Strachan}}, \binits{L.}},
\bauthor{\bsnm{{von Steiger}}, \binits{R.}},
\bauthor{\bsnm{{Wang}}, \binits{Y.-M.}}:
\byear{2016},
\batitle{{Slow Solar Wind: Observations and Modeling}}.
\bjtitle{\ssr}
\bvolume{201},
\bfpage{55}.
\doiurl{https://doi.org/10.1007/s11214-016-0264-1}.
\adsurl{2016SSRv..201...55A}.
\end{barticle}
\endbibitem

\bibitem[\protect\citeauthoryear{{Antiochos}, {DeVore}, and
  {Klimchuk}}{1999}]{antiochos99}
\begin{barticle}
\bauthor{\bsnm{{Antiochos}}, \binits{S.K.}},
\bauthor{\bsnm{{DeVore}}, \binits{C.R.}},
\bauthor{\bsnm{{Klimchuk}}, \binits{J.A.}}:
\byear{1999},
\batitle{{A Model for Solar Coronal Mass Ejections}}.
\bjtitle{\apj}
\bvolume{510},
\bfpage{485}.
\doiurl{https://doi.org/10.1086/306563}.
\adsurl{1999ApJ...510..485A}.
\end{barticle}
\endbibitem

\bibitem[\protect\citeauthoryear{{Asplund} et~al.}{2009}]{asplund09}
\begin{barticle}
\bauthor{\bsnm{{Asplund}}, \binits{M.}},
\bauthor{\bsnm{{Grevesse}}, \binits{N.}},
\bauthor{\bsnm{{Sauval}}, \binits{A.J.}},
\bauthor{\bsnm{{Scott}}, \binits{P.}}:
\byear{2009},
\batitle{{The Chemical Composition of the Sun}}.
\bjtitle{\araa}
\bvolume{47},
\bfpage{481}.
\doiurl{https://doi.org/10.1146/annurev.astro.46.060407.145222}.
\adsurl{2009ARA&A..47..481A}.
\end{barticle}
\endbibitem

\bibitem[\protect\citeauthoryear{{Burlaga} et~al.}{1981}]{burlaga81}
\begin{barticle}
\bauthor{\bsnm{{Burlaga}}, \binits{L.}},
\bauthor{\bsnm{{Sittler}}, \binits{E.}},
\bauthor{\bsnm{{Mariani}}, \binits{F.}},
\bauthor{\bsnm{{Schwenn}}, \binits{R.}}:
\byear{1981},
\batitle{{Magnetic loop behind an interplanetary shock: Voyager, Helios, and
  IMP 8 observations}}.
\bjtitle{\jgr}
\bvolume{86},
\bfpage{6673}.
\doiurl{https://doi.org/10.1029/JA086iA08p06673}.
\adsurl{1981JGR....86.6673B}.
\end{barticle}
\endbibitem

\bibitem[\protect\citeauthoryear{{Caffau} et~al.}{2011}]{caffau11}
\begin{barticle}
\bauthor{\bsnm{{Caffau}}, \binits{E.}},
\bauthor{\bsnm{{Ludwig}}, \binits{H.-G.}},
\bauthor{\bsnm{{Steffen}}, \binits{M.}},
\bauthor{\bsnm{{Freytag}}, \binits{B.}},
\bauthor{\bsnm{{Bonifacio}}, \binits{P.}}:
\byear{2011},
\batitle{{Solar Chemical Abundances Determined with a CO5BOLD 3D Model
  Atmosphere}}.
\bjtitle{\solphys}
\bvolume{268},
\bfpage{255}.
\doiurl{https://doi.org/10.1007/s11207-010-9541-4}.
\adsurl{2011SoPh..268..255C}.
\end{barticle}
\endbibitem

\bibitem[\protect\citeauthoryear{{Charbonneau}}{2020}]{charbonneau20}
\begin{barticle}
\bauthor{\bsnm{{Charbonneau}}, \binits{P.}}:
\byear{2020},
\batitle{{Dynamo models of the solar cycle}}.
\bjtitle{Living Reviews in Solar Physics}
\bvolume{17},
\bfpage{4}.
\doiurl{https://doi.org/10.1007/s41116-020-00025-6}.
\adsurl{2020LRSP...17....4C}.
\end{barticle}
\endbibitem

\bibitem[\protect\citeauthoryear{{Chen}}{2008}]{chenpengfei08}
\begin{barticle}
\bauthor{\bsnm{{Chen}}, \binits{P.F.}}:
\byear{2008},
\batitle{{Initiation and propagation of coronal mass ejections}}.
\bjtitle{Journal of Astrophysics and Astronomy}
\bvolume{29},
\bfpage{179}.
\doiurl{https://doi.org/10.1007/s12036-008-0023-0}.
\adsurl{2008JApA...29..179C}.
\end{barticle}
\endbibitem

\bibitem[\protect\citeauthoryear{{Chen}}{2011}]{chenpengfei11}
\begin{barticle}
\bauthor{\bsnm{{Chen}}, \binits{P.F.}}:
\byear{2011},
\batitle{{Coronal Mass Ejections: Models and Their Observational Basis}}.
\bjtitle{Living Reviews in Solar Physics}
\bvolume{8},
\bfpage{1}.
\doiurl{https://doi.org/10.12942/lrsp-2011-1}.
\adsurl{2011LRSP....8....1C}.
\end{barticle}
\endbibitem

\bibitem[\protect\citeauthoryear{{Chen}, {Esser}, and {Hu}}{2003}]{chenyao03}
\begin{barticle}
\bauthor{\bsnm{{Chen}}, \binits{Y.}},
\bauthor{\bsnm{{Esser}}, \binits{R.}},
\bauthor{\bsnm{{Hu}}, \binits{Y.}}:
\byear{2003},
\batitle{{Formation of Minor-Ion Charge States in the Fast Solar Wind: Roles of
  Differential Flow Speeds of Ions of the Same Element}}.
\bjtitle{\apj}
\bvolume{582},
\bfpage{467}.
\doiurl{https://doi.org/10.1086/344642}.
\adsurl{2003ApJ...582..467C}.
\end{barticle}
\endbibitem

\bibitem[\protect\citeauthoryear{{Chen}, {Hu}, and {Sun}}{2007}]{chenyao07a}
\begin{barticle}
\bauthor{\bsnm{{Chen}}, \binits{Y.}},
\bauthor{\bsnm{{Hu}}, \binits{Y.Q.}},
\bauthor{\bsnm{{Sun}}, \binits{S.J.}}:
\byear{2007},
\batitle{{Catastrophic Eruption of Magnetic Flux Rope in the Corona and Solar
  Wind With and Without Magnetic Reconnection}}.
\bjtitle{\apj}
\bvolume{665},
\bfpage{1421}.
\doiurl{https://doi.org/10.1086/519551}.
\adsurl{2007ApJ...665.1421C}.
\end{barticle}
\endbibitem

\bibitem[\protect\citeauthoryear{{Cheng} et~al.}{2014}]{chengxin14a}
\begin{barticle}
\bauthor{\bsnm{{Cheng}}, \binits{X.}},
\bauthor{\bsnm{{Ding}}, \binits{M.D.}},
\bauthor{\bsnm{{Zhang}}, \binits{J.}},
\bauthor{\bsnm{{Srivastava}}, \binits{A.K.}},
\bauthor{\bsnm{{Guo}}, \binits{Y.}},
\bauthor{\bsnm{{Chen}}, \binits{P.F.}},
\bauthor{\bsnm{{Sun}}, \binits{J.Q.}}:
\byear{2014},
\batitle{{On the Relationship Between a Hot-channel-like Solar Magnetic Flux
  Rope and its Embedded Prominence}}.
\bjtitle{\apjl}
\bvolume{789},
\bfpage{L35}.
\doiurl{https://doi.org/10.1088/2041-8205/789/2/L35}.
\adsurl{2014ApJ...789L..35C}.
\end{barticle}
\endbibitem

\bibitem[\protect\citeauthoryear{{Cheng} et~al.}{2020}]{chengxin20}
\begin{barticle}
\bauthor{\bsnm{{Cheng}}, \binits{X.}},
\bauthor{\bsnm{{Zhang}}, \binits{J.}},
\bauthor{\bsnm{{Kliem}}, \binits{B.}},
\bauthor{\bsnm{{T{\"o}r{\"o}k}}, \binits{T.}},
\bauthor{\bsnm{{Xing}}, \binits{C.}},
\bauthor{\bsnm{{Zhou}}, \binits{Z.J.}},
\bauthor{\bsnm{{Inhester}}, \binits{B.}},
\bauthor{\bsnm{{Ding}}, \binits{M.D.}}:
\byear{2020},
\batitle{{Initiation and Early Kinematic Evolution of Solar Eruptions}}.
\bjtitle{\apj}
\bvolume{894},
\bfpage{85}.
\doiurl{https://doi.org/10.3847/1538-4357/ab886a}.
\adsurl{2020ApJ...894...85C}.
\end{barticle}
\endbibitem

\bibitem[\protect\citeauthoryear{{Chi} et~al.}{2016}]{chiyutian16}
\begin{barticle}
\bauthor{\bsnm{{Chi}}, \binits{Y.}},
\bauthor{\bsnm{{Shen}}, \binits{C.}},
\bauthor{\bsnm{{Wang}}, \binits{Y.}},
\bauthor{\bsnm{{Xu}}, \binits{M.}},
\bauthor{\bsnm{{Ye}}, \binits{P.}},
\bauthor{\bsnm{{Wang}}, \binits{S.}}:
\byear{2016},
\batitle{{Statistical Study of the Interplanetary Coronal Mass Ejections from
  1995 to 2015}}.
\bjtitle{\solphys}
\bvolume{291},
\bfpage{2419}.
\doiurl{https://doi.org/10.1007/s11207-016-0971-5}.
\adsurl{2016SoPh..291.2419C}.
\end{barticle}
\endbibitem

\bibitem[\protect\citeauthoryear{{Cranmer} and {Winebarger}}{2019}]{cranmer19}
\begin{barticle}
\bauthor{\bsnm{{Cranmer}}, \binits{S.R.}},
\bauthor{\bsnm{{Winebarger}}, \binits{A.R.}}:
\byear{2019},
\batitle{{The Properties of the Solar Corona and Its Connection to the Solar
  Wind}}.
\bjtitle{\araa}
\bvolume{57},
\bfpage{157}.
\doiurl{https://doi.org/10.1146/annurev-astro-091918-104416}.
\adsurl{2019ARA&A..57..157C}.
\end{barticle}
\endbibitem

\bibitem[\protect\citeauthoryear{{Del Zanna} and {Mason}}{2018}]{DelZanna18}
\begin{barticle}
\bauthor{\bsnm{{Del Zanna}}, \binits{G.}},
\bauthor{\bsnm{{Mason}}, \binits{H.E.}}:
\byear{2018},
\batitle{{Solar UV and X-ray spectral diagnostics}}.
\bjtitle{Living Reviews in Solar Physics}
\bvolume{15},
\bfpage{5}.
\doiurl{https://doi.org/10.1007/s41116-018-0015-3}.
\adsurl{2018LRSP...15....5D}.
\end{barticle}
\endbibitem

\bibitem[\protect\citeauthoryear{{Dikpati}, {Suresh}, and
  {Burkepile}}{2016}]{dikpati16}
\begin{barticle}
\bauthor{\bsnm{{Dikpati}}, \binits{M.}},
\bauthor{\bsnm{{Suresh}}, \binits{A.}},
\bauthor{\bsnm{{Burkepile}}, \binits{J.}}:
\byear{2016},
\batitle{{Cyclic Evolution of Coronal Fields from a Coupled Dynamo
  Potential-Field Source-Surface Model}}.
\bjtitle{\solphys}
\bvolume{291},
\bfpage{339}.
\doiurl{https://doi.org/10.1007/s11207-015-0831-8}.
\adsurl{2016SoPh..291..339D}.
\end{barticle}
\endbibitem

\bibitem[\protect\citeauthoryear{{Fan} and {Gibson}}{2007}]{fanyuhong07}
\begin{barticle}
\bauthor{\bsnm{{Fan}}, \binits{Y.}},
\bauthor{\bsnm{{Gibson}}, \binits{S.E.}}:
\byear{2007},
\batitle{{Onset of Coronal Mass Ejections Due to Loss of Confinement of Coronal
  Flux Ropes}}.
\bjtitle{\apj}
\bvolume{668},
\bfpage{1232}.
\doiurl{https://doi.org/10.1086/521335}.
\adsurl{2007ApJ...668.1232F}.
\end{barticle}
\endbibitem

\bibitem[\protect\citeauthoryear{{Feldman} et~al.}{1998}]{feldman98}
\begin{barticle}
\bauthor{\bsnm{{Feldman}}, \binits{U.}},
\bauthor{\bsnm{{Sch{\"u}hle}}, \binits{U.}},
\bauthor{\bsnm{{Widing}}, \binits{K.G.}},
\bauthor{\bsnm{{Laming}}, \binits{J.M.}}:
\byear{1998},
\batitle{{Coronal Composition above the Solar Equator and the North Pole as
  Determined from Spectra Acquired by the SUMER Instrument on SOHO}}.
\bjtitle{\apj}
\bvolume{505},
\bfpage{999}.
\doiurl{https://doi.org/10.1086/306195}.
\adsurl{1998ApJ...505..999F}.
\end{barticle}
\endbibitem

\bibitem[\protect\citeauthoryear{{Feng} et~al.}{2018}]{fengxuedong18}
\begin{barticle}
\bauthor{\bsnm{{Feng}}, \binits{X.}},
\bauthor{\bsnm{{Yao}}, \binits{S.}},
\bauthor{\bsnm{{Li}}, \binits{D.}},
\bauthor{\bsnm{{Li}}, \binits{G.}},
\bauthor{\bsnm{{Yan}}, \binits{X.}}:
\byear{2018},
\batitle{{Statistical Study of ICMEs with Low Mean Carbon Charge State Plasmas
  Detected from 1998 to 2011}}.
\bjtitle{\apj}
\bvolume{868},
\bfpage{124}.
\doiurl{https://doi.org/10.3847/1538-4357/aae92c}.
\adsurl{2018ApJ...868..124F}.
\end{barticle}
\endbibitem

\bibitem[\protect\citeauthoryear{{Forbes}}{2000}]{forbes00}
\begin{barticle}
\bauthor{\bsnm{{Forbes}}, \binits{T.G.}}:
\byear{2000},
\batitle{{A review on the genesis of coronal mass ejections}}.
\bjtitle{\jgr}
\bvolume{105},
\bfpage{23153}.
\doiurl{https://doi.org/10.1029/2000JA000005}.
\adsurl{2000JGR...10523153F}.
\end{barticle}
\endbibitem

\bibitem[\protect\citeauthoryear{{Forbes} and {Priest}}{1995}]{forbes95}
\begin{barticle}
\bauthor{\bsnm{{Forbes}}, \binits{T.G.}},
\bauthor{\bsnm{{Priest}}, \binits{E.R.}}:
\byear{1995},
\batitle{{Photospheric Magnetic Field Evolution and Eruptive Flares}}.
\bjtitle{\apj}
\bvolume{446},
\bfpage{377}.
\doiurl{https://doi.org/10.1086/175797}.
\adsurl{1995ApJ...446..377F}.
\end{barticle}
\endbibitem

\bibitem[\protect\citeauthoryear{{Fu} et~al.}{2020}]{fuhui20}
\begin{barticle}
\bauthor{\bsnm{{Fu}}, \binits{H.}},
\bauthor{\bsnm{{Harrison}}, \binits{R.A.}},
\bauthor{\bsnm{{Davies}}, \binits{J.A.}},
\bauthor{\bsnm{{Xia}}, \binits{L.}},
\bauthor{\bsnm{{Zhu}}, \binits{X.}},
\bauthor{\bsnm{{Li}}, \binits{B.}},
\bauthor{\bsnm{{Huang}}, \binits{Z.}},
\bauthor{\bsnm{{Barnes}}, \binits{D.}}:
\byear{2020},
\batitle{{The High Helium Abundance and Charge States of the Interplanetary CME
  and Its Material Source on the Sun}}.
\bjtitle{\apjl}
\bvolume{900},
\bfpage{L18}.
\doiurl{https://doi.org/10.3847/2041-8213/abb083}.
\adsurl{2020ApJ...900L..18F}.
\end{barticle}
\endbibitem

\bibitem[\protect\citeauthoryear{{Gilbert} et~al.}{2012}]{gilbert12}
\begin{barticle}
\bauthor{\bsnm{{Gilbert}}, \binits{J.A.}},
\bauthor{\bsnm{{Lepri}}, \binits{S.T.}},
\bauthor{\bsnm{{Landi}}, \binits{E.}},
\bauthor{\bsnm{{Zurbuchen}}, \binits{T.H.}}:
\byear{2012},
\batitle{{First Measurements of the Complete Heavy-ion Charge State
  Distributions of C, O, and Fe Associated with Interplanetary Coronal Mass
  Ejections}}.
\bjtitle{\apj}
\bvolume{751},
\bfpage{20}.
\doiurl{https://doi.org/10.1088/0004-637X/751/1/20}.
\adsurl{2012ApJ...751...20G}.
\end{barticle}
\endbibitem

\bibitem[\protect\citeauthoryear{{Gloeckler} et~al.}{1998}]{gloeckler98}
\begin{barticle}
\bauthor{\bsnm{{Gloeckler}}, \binits{G.}},
\bauthor{\bsnm{{Cain}}, \binits{J.}},
\bauthor{\bsnm{{Ipavich}}, \binits{F.M.}},
\bauthor{\bsnm{{Tums}}, \binits{E.O.}},
\bauthor{\bsnm{{Bedini}}, \binits{P.}},
\bauthor{\bsnm{{Fisk}}, \binits{L.A.}},
\bauthor{\bsnm{{Zurbuchen}}, \binits{T.H.}},
\bauthor{\bsnm{{Bochsler}}, \binits{P.}},
\bauthor{\bsnm{{Fischer}}, \binits{J.}},
\bauthor{\bsnm{{Wimmer-Schweingruber}}, \binits{R.F.}},
\bauthor{\bsnm{{Geiss}}, \binits{J.}},
\bauthor{\bsnm{{Kallenbach}}, \binits{R.}}:
\byear{1998},
\batitle{{Investigation of the composition of solar and interstellar matter
  using solar wind and pickup ion measurements with SWICS and SWIMS on the ACE
  spacecraft}}.
\bjtitle{\ssr}
\bvolume{86},
\bfpage{497}.
\doiurl{https://doi.org/10.1023/A:1005036131689}.
\adsurl{1998SSRv...86..497G}.
\end{barticle}
\endbibitem

\bibitem[\protect\citeauthoryear{{Gosling} et~al.}{1991}]{gosling91}
\begin{barticle}
\bauthor{\bsnm{{Gosling}}, \binits{J.T.}},
\bauthor{\bsnm{{McComas}}, \binits{D.J.}},
\bauthor{\bsnm{{Phillips}}, \binits{J.L.}},
\bauthor{\bsnm{{Bame}}, \binits{S.J.}}:
\byear{1991},
\batitle{{Geomagnetic activity associated with earth passage of interplanetary
  shock disturbances and coronal mass ejections}}.
\bjtitle{\jgr}
\bvolume{96},
\bfpage{7831}.
\doiurl{https://doi.org/10.1029/91JA00316}.
\adsurl{1991JGR....96.7831G}.
\end{barticle}
\endbibitem

\bibitem[\protect\citeauthoryear{{Grevesse} and {Sauval}}{1998}]{grevesse98}
\begin{barticle}
\bauthor{\bsnm{{Grevesse}}, \binits{N.}},
\bauthor{\bsnm{{Sauval}}, \binits{A.J.}}:
\byear{1998},
\batitle{{Standard Solar Composition}}.
\bjtitle{\ssr}
\bvolume{85},
\bfpage{161}.
\doiurl{https://doi.org/10.1023/A:1005161325181}.
\adsurl{1998SSRv...85..161G}.
\end{barticle}
\endbibitem

\bibitem[\protect\citeauthoryear{{Gruesbeck}, {Lepri}, and
  {Zurbuchen}}{2012}]{gruesbeck12}
\begin{barticle}
\bauthor{\bsnm{{Gruesbeck}}, \binits{J.R.}},
\bauthor{\bsnm{{Lepri}}, \binits{S.T.}},
\bauthor{\bsnm{{Zurbuchen}}, \binits{T.H.}}:
\byear{2012},
\batitle{{Two-plasma Model for Low Charge State Interplanetary Coronal Mass
  Ejection Observations}}.
\bjtitle{\apj}
\bvolume{760},
\bfpage{141}.
\doiurl{https://doi.org/10.1088/0004-637X/760/2/141}.
\adsurl{2012ApJ...760..141G}.
\end{barticle}
\endbibitem

\bibitem[\protect\citeauthoryear{{Gruesbeck} et~al.}{2011}]{gruesbeck11}
\begin{barticle}
\bauthor{\bsnm{{Gruesbeck}}, \binits{J.R.}},
\bauthor{\bsnm{{Lepri}}, \binits{S.T.}},
\bauthor{\bsnm{{Zurbuchen}}, \binits{T.H.}},
\bauthor{\bsnm{{Antiochos}}, \binits{S.K.}}:
\byear{2011},
\batitle{{Constraints on Coronal Mass Ejection Evolution from in Situ
  Observations of Ionic Charge States}}.
\bjtitle{\apj}
\bvolume{730},
\bfpage{103}.
\doiurl{https://doi.org/10.1088/0004-637X/730/2/103}.
\adsurl{2011ApJ...730..103G}.
\end{barticle}
\endbibitem

\bibitem[\protect\citeauthoryear{{Gu}, {Yao}, and {Dai}}{2020}]{guchaoran20}
\begin{barticle}
\bauthor{\bsnm{{Gu}}, \binits{C.}},
\bauthor{\bsnm{{Yao}}, \binits{S.}},
\bauthor{\bsnm{{Dai}}, \binits{L.}}:
\byear{2020},
\batitle{{Abundances and Charge States of Heavy Ions in ICMEs Highly Related to
  Speed and Solar Activity}}.
\bjtitle{\apj}
\bvolume{900},
\bfpage{123}.
\doiurl{https://doi.org/10.3847/1538-4357/aba7b8}.
\adsurl{2020ApJ...900..123G}.
\end{barticle}
\endbibitem

\bibitem[\protect\citeauthoryear{{Huang} et~al.}{2018}]{huangjia18}
\begin{barticle}
\bauthor{\bsnm{{Huang}}, \binits{J.}},
\bauthor{\bsnm{{Liu}}, \binits{Y.C.-M.}},
\bauthor{\bsnm{{Peng}}, \binits{J.}},
\bauthor{\bsnm{{Qi}}, \binits{Z.}},
\bauthor{\bsnm{{Li}}, \binits{H.}},
\bauthor{\bsnm{{Klecker}}, \binits{B.}},
\bauthor{\bsnm{{Song}}, \binits{H.}},
\bauthor{\bsnm{{Zheng}}, \binits{J.}},
\bauthor{\bsnm{{Hu}}, \binits{Q.}}:
\byear{2018},
\batitle{{The Distributions of Iron Average Charge States in Small Flux Ropes
  in Interplanetary Space: Clues to Their Twisted Structures}}.
\bjtitle{Journal of Geophysical Research (Space Physics)}
\bvolume{123},
\bfpage{7167}.
\doiurl{https://doi.org/10.1029/2018JA025660}.
\adsurl{2018JGRA..123.7167H}.
\end{barticle}
\endbibitem

\bibitem[\protect\citeauthoryear{{Huang} et~al.}{2020}]{huangjin20}
\begin{barticle}
\bauthor{\bsnm{{Huang}}, \binits{J.}},
\bauthor{\bsnm{{Liu}}, \binits{Y.}},
\bauthor{\bsnm{{Feng}}, \binits{H.}},
\bauthor{\bsnm{{Zhao}}, \binits{A.}},
\bauthor{\bsnm{{Abidin}}, \binits{Z.Z.}},
\bauthor{\bsnm{{Shen}}, \binits{Y.}},
\bauthor{\bsnm{{Jacob}}, \binits{O.}}:
\byear{2020},
\batitle{{A Statistical Study of the Plasma and Composition Distribution inside
  Magnetic Clouds: 1998-2011}}.
\bjtitle{\apj}
\bvolume{893},
\bfpage{136}.
\doiurl{https://doi.org/10.3847/1538-4357/ab7a28}.
\adsurl{2020ApJ...893..136H}.
\end{barticle}
\endbibitem

\bibitem[\protect\citeauthoryear{{Jian} et~al.}{2018}]{jianlan18}
\begin{barticle}
\bauthor{\bsnm{{Jian}}, \binits{L.K.}},
\bauthor{\bsnm{{Russell}}, \binits{C.T.}},
\bauthor{\bsnm{{Luhmann}}, \binits{J.G.}},
\bauthor{\bsnm{{Galvin}}, \binits{A.B.}}:
\byear{2018},
\batitle{{STEREO Observations of Interplanetary Coronal Mass Ejections in
  2007-2016}}.
\bjtitle{\apj}
\bvolume{855},
\bfpage{114}.
\doiurl{https://doi.org/10.3847/1538-4357/aab189}.
\adsurl{2018ApJ...855..114J}.
\end{barticle}
\endbibitem

\bibitem[\protect\citeauthoryear{{Kocher} et~al.}{2017}]{kocher17}
\begin{barticle}
\bauthor{\bsnm{{Kocher}}, \binits{M.}},
\bauthor{\bsnm{{Lepri}}, \binits{S.T.}},
\bauthor{\bsnm{{Landi}}, \binits{E.}},
\bauthor{\bsnm{{Zhao}}, \binits{L.}},
\bauthor{\bsnm{{Manchester}}, \binits{I.} \bsuffix{W.~B.}}:
\byear{2017},
\batitle{{Anatomy of Depleted Interplanetary Coronal Mass Ejections}}.
\bjtitle{\apj}
\bvolume{834},
\bfpage{147}.
\doiurl{https://doi.org/10.3847/1538-4357/834/2/147}.
\adsurl{2017ApJ...834..147K}.
\end{barticle}
\endbibitem

\bibitem[\protect\citeauthoryear{{Kou} et~al.}{2020}]{kouyuankun20}
\begin{barticle}
\bauthor{\bsnm{{Kou}}, \binits{Y.K.}},
\bauthor{\bsnm{{Jing}}, \binits{Z.C.}},
\bauthor{\bsnm{{Cheng}}, \binits{X.}},
\bauthor{\bsnm{{Pan}}, \binits{W.Q.}},
\bauthor{\bsnm{{Liu}}, \binits{Y.}},
\bauthor{\bsnm{{Li}}, \binits{C.}},
\bauthor{\bsnm{{Ding}}, \binits{M.D.}}:
\byear{2020},
\batitle{{What Determines Solar Flares Producing Interplanetary Type III Radio
  Bursts?}}
\bjtitle{\apjl}
\bvolume{898},
\bfpage{L24}.
\doiurl{https://doi.org/10.3847/2041-8213/aba362}.
\adsurl{2020ApJ...898L..24K}.
\end{barticle}
\endbibitem

\bibitem[\protect\citeauthoryear{{Laming}}{2004}]{laming04}
\begin{barticle}
\bauthor{\bsnm{{Laming}}, \binits{J.M.}}:
\byear{2004},
\batitle{{A Unified Picture of the First Ionization Potential and Inverse First
  Ionization Potential Effects}}.
\bjtitle{\apj}
\bvolume{614},
\bfpage{1063}.
\doiurl{https://doi.org/10.1086/423780}.
\adsurl{2004ApJ...614.1063L}.
\end{barticle}
\endbibitem

\bibitem[\protect\citeauthoryear{{Laming}}{2012}]{laming12}
\begin{barticle}
\bauthor{\bsnm{{Laming}}, \binits{J.M.}}:
\byear{2012},
\batitle{{Non-WKB Models of the First Ionization Potential Effect: The Role of
  Slow Mode Waves}}.
\bjtitle{\apj}
\bvolume{744},
\bfpage{115}.
\doiurl{https://doi.org/10.1088/0004-637X/744/2/115}.
\adsurl{2012ApJ...744..115L}.
\end{barticle}
\endbibitem

\bibitem[\protect\citeauthoryear{{Laming}}{2015}]{laming15}
\begin{barticle}
\bauthor{\bsnm{{Laming}}, \binits{J.M.}}:
\byear{2015},
\batitle{{The FIP and Inverse FIP Effects in Solar and Stellar Coronae}}.
\bjtitle{Living Reviews in Solar Physics}
\bvolume{12},
\bfpage{2}.
\doiurl{https://doi.org/10.1007/lrsp-2015-2}.
\adsurl{2015LRSP...12....2L}.
\end{barticle}
\endbibitem

\bibitem[\protect\citeauthoryear{{Landi} et~al.}{2012a}]{landi12a}
\begin{barticle}
\bauthor{\bsnm{{Landi}}, \binits{E.}},
\bauthor{\bsnm{{Alexander}}, \binits{R.L.}},
\bauthor{\bsnm{{Gruesbeck}}, \binits{J.R.}},
\bauthor{\bsnm{{Gilbert}}, \binits{J.A.}},
\bauthor{\bsnm{{Lepri}}, \binits{S.T.}},
\bauthor{\bsnm{{Manchester}}, \binits{W.B.}},
\bauthor{\bsnm{{Zurbuchen}}, \binits{T.H.}}:
\byear{2012}a,
\batitle{{Carbon Ionization Stages as a Diagnostic of the Solar Wind}}.
\bjtitle{\apj}
\bvolume{744},
\bfpage{100}.
\doiurl{https://doi.org/10.1088/0004-637X/744/2/100}.
\adsurl{2012ApJ...744..100L}.
\end{barticle}
\endbibitem

\bibitem[\protect\citeauthoryear{{Landi} et~al.}{2012b}]{landi12b}
\begin{barticle}
\bauthor{\bsnm{{Landi}}, \binits{E.}},
\bauthor{\bsnm{{Gruesbeck}}, \binits{J.R.}},
\bauthor{\bsnm{{Lepri}}, \binits{S.T.}},
\bauthor{\bsnm{{Zurbuchen}}, \binits{T.H.}}:
\byear{2012}b,
\batitle{{New Solar Wind Diagnostic Using Both in Situ and Spectroscopic
  Measurements}}.
\bjtitle{\apj}
\bvolume{750},
\bfpage{159}.
\doiurl{https://doi.org/10.1088/0004-637X/750/2/159}.
\adsurl{2012ApJ...750..159L}.
\end{barticle}
\endbibitem

\bibitem[\protect\citeauthoryear{{Lepri} and {Zurbuchen}}{2004}]{lepri04}
\begin{barticle}
\bauthor{\bsnm{{Lepri}}, \binits{S.T.}},
\bauthor{\bsnm{{Zurbuchen}}, \binits{T.H.}}:
\byear{2004},
\batitle{{Iron charge state distributions as an indicator of hot ICMEs:
  Possible sources and temporal and spatial variations during solar maximum}}.
\bjtitle{Journal of Geophysical Research (Space Physics)}
\bvolume{109},
\bfpage{A01112}.
\doiurl{https://doi.org/10.1029/2003JA009954}.
\adsurl{2004JGRA..109.1112L}.
\end{barticle}
\endbibitem

\bibitem[\protect\citeauthoryear{{Lepri} and {Zurbuchen}}{2010}]{lepri10}
\begin{barticle}
\bauthor{\bsnm{{Lepri}}, \binits{S.T.}},
\bauthor{\bsnm{{Zurbuchen}}, \binits{T.H.}}:
\byear{2010},
\batitle{{Direct Observational Evidence of Filament Material Within
  Interplanetary Coronal Mass Ejections}}.
\bjtitle{\apjl}
\bvolume{723},
\bfpage{L22}.
\doiurl{https://doi.org/10.1088/2041-8205/723/1/L22}.
\adsurl{2010ApJ...723L..22L}.
\end{barticle}
\endbibitem

\bibitem[\protect\citeauthoryear{{Lepri}, {Landi}, and
  {Zurbuchen}}{2013}]{lepri13}
\begin{barticle}
\bauthor{\bsnm{{Lepri}}, \binits{S.T.}},
\bauthor{\bsnm{{Landi}}, \binits{E.}},
\bauthor{\bsnm{{Zurbuchen}}, \binits{T.H.}}:
\byear{2013},
\batitle{{Solar Wind Heavy Ions over Solar Cycle 23: ACE/SWICS Measurements}}.
\bjtitle{\apj}
\bvolume{768},
\bfpage{94}.
\doiurl{https://doi.org/10.1088/0004-637X/768/1/94}.
\adsurl{2013ApJ...768...94L}.
\end{barticle}
\endbibitem

\bibitem[\protect\citeauthoryear{{Lepri} et~al.}{2001}]{lepri01}
\begin{barticle}
\bauthor{\bsnm{{Lepri}}, \binits{S.T.}},
\bauthor{\bsnm{{Zurbuchen}}, \binits{T.H.}},
\bauthor{\bsnm{{Fisk}}, \binits{L.A.}},
\bauthor{\bsnm{{Richardson}}, \binits{I.G.}},
\bauthor{\bsnm{{Cane}}, \binits{H.V.}},
\bauthor{\bsnm{{Gloeckler}}, \binits{G.}}:
\byear{2001},
\batitle{{Iron charge distribution as an identifier of interplanetary coronal
  mass ejections}}.
\bjtitle{\jgr}
\bvolume{106},
\bfpage{29231}.
\doiurl{https://doi.org/10.1029/2001JA000014}.
\adsurl{2001JGR...10629231L}.
\end{barticle}
\endbibitem

\bibitem[\protect\citeauthoryear{{Li} et~al.}{2020}]{libo20}
\begin{barticle}
\bauthor{\bsnm{{Li}}, \binits{B.}},
\bauthor{\bsnm{{Antolin}}, \binits{P.}},
\bauthor{\bsnm{{Guo}}, \binits{M.-Z.}},
\bauthor{\bsnm{{Kuznetsov}}, \binits{A.A.}},
\bauthor{\bsnm{{Pascoe}}, \binits{D.J.}},
\bauthor{\bsnm{{Van Doorsselaere}}, \binits{T.}},
\bauthor{\bsnm{{Vasheghani Farahani}}, \binits{S.}}:
\byear{2020},
\batitle{{Magnetohydrodynamic Fast Sausage Waves in the Solar Corona}}.
\bjtitle{\ssr}
\bvolume{216},
\bfpage{136}.
\doiurl{https://doi.org/10.1007/s11214-020-00761-z}.
\adsurl{2020SSRv..216..136L}.
\end{barticle}
\endbibitem

\bibitem[\protect\citeauthoryear{{Lin} and {Forbes}}{2000}]{linjun00}
\begin{barticle}
\bauthor{\bsnm{{Lin}}, \binits{J.}},
\bauthor{\bsnm{{Forbes}}, \binits{T.G.}}:
\byear{2000},
\batitle{{Effects of reconnection on the coronal mass ejection process}}.
\bjtitle{\jgr}
\bvolume{105},
\bfpage{2375}.
\doiurl{https://doi.org/10.1029/1999JA900477}.
\adsurl{2000JGR...105.2375L}.
\end{barticle}
\endbibitem

\bibitem[\protect\citeauthoryear{{Liu} et~al.}{2014}]{liuying14}
\begin{barticle}
\bauthor{\bsnm{{Liu}}, \binits{Y.D.}},
\bauthor{\bsnm{{Luhmann}}, \binits{J.G.}},
\bauthor{\bsnm{{Kajdi{\v{c}}}}, \binits{P.}},
\bauthor{\bsnm{{Kilpua}}, \binits{E.K.J.}},
\bauthor{\bsnm{{Lugaz}}, \binits{N.}},
\bauthor{\bsnm{{Nitta}}, \binits{N.V.}},
\bauthor{\bsnm{{M{\"o}stl}}, \binits{C.}},
\bauthor{\bsnm{{Lavraud}}, \binits{B.}},
\bauthor{\bsnm{{Bale}}, \binits{S.D.}},
\bauthor{\bsnm{{Farrugia}}, \binits{C.J.}},
\bauthor{\bsnm{{Galvin}}, \binits{A.B.}}:
\byear{2014},
\batitle{{Observations of an extreme storm in interplanetary space caused by
  successive coronal mass ejections}}.
\bjtitle{Nature Communications}
\bvolume{5},
\bfpage{3481}.
\doiurl{https://doi.org/10.1038/ncomms4481}.
\adsurl{2014NatCo...5.3481L}.
\end{barticle}
\endbibitem

\bibitem[\protect\citeauthoryear{{Lynch} et~al.}{2011}]{lynch11}
\begin{barticle}
\bauthor{\bsnm{{Lynch}}, \binits{B.J.}},
\bauthor{\bsnm{{Reinard}}, \binits{A.A.}},
\bauthor{\bsnm{{Mulligan}}, \binits{T.}},
\bauthor{\bsnm{{Reeves}}, \binits{K.K.}},
\bauthor{\bsnm{{Rakowski}}, \binits{C.E.}},
\bauthor{\bsnm{{Allred}}, \binits{J.C.}},
\bauthor{\bsnm{{Li}}, \binits{Y.}},
\bauthor{\bsnm{{Laming}}, \binits{J.M.}},
\bauthor{\bsnm{{MacNeice}}, \binits{P.J.}},
\bauthor{\bsnm{{Linker}}, \binits{J.A.}}:
\byear{2011},
\batitle{{Ionic Composition Structure of Coronal Mass Ejections in Axisymmetric
  Magnetohydrodynamic Models}}.
\bjtitle{\apj}
\bvolume{740},
\bfpage{112}.
\doiurl{https://doi.org/10.1088/0004-637X/740/2/112}.
\adsurl{2011ApJ...740..112L}.
\end{barticle}
\endbibitem

\bibitem[\protect\citeauthoryear{{Manchester} et~al.}{2017}]{manchester17}
\begin{barticle}
\bauthor{\bsnm{{Manchester}}, \binits{W.}},
\bauthor{\bsnm{{Kilpua}}, \binits{E.K.J.}},
\bauthor{\bsnm{{Liu}}, \binits{Y.D.}},
\bauthor{\bsnm{{Lugaz}}, \binits{N.}},
\bauthor{\bsnm{{Riley}}, \binits{P.}},
\bauthor{\bsnm{{T{\"o}r{\"o}k}}, \binits{T.}},
\bauthor{\bsnm{{Vr{\v{s}}nak}}, \binits{B.}}:
\byear{2017},
\batitle{{The Physical Processes of CME/ICME Evolution}}.
\bjtitle{\ssr}
\bvolume{212},
\bfpage{1159}.
\doiurl{https://doi.org/10.1007/s11214-017-0394-0}.
\adsurl{2017SSRv..212.1159M}.
\end{barticle}
\endbibitem

\bibitem[\protect\citeauthoryear{{Mari{\v{c}}i{\'c}} et~al.}{2007}]{maricic07}
\begin{barticle}
\bauthor{\bsnm{{Mari{\v{c}}i{\'c}}}, \binits{D.}},
\bauthor{\bsnm{{Vr{\v{s}}nak}}, \binits{B.}},
\bauthor{\bsnm{{Stanger}}, \binits{A.L.}},
\bauthor{\bsnm{{Veronig}}, \binits{A.M.}},
\bauthor{\bsnm{{Temmer}}, \binits{M.}},
\bauthor{\bsnm{{Ro{\v{s}}a}}, \binits{D.}}:
\byear{2007},
\batitle{{Acceleration Phase of Coronal Mass Ejections: II. Synchronization of
  the Energy Release in the Associated Flare}}.
\bjtitle{\solphys}
\bvolume{241},
\bfpage{99}.
\doiurl{https://doi.org/10.1007/s11207-007-0291-x}.
\adsurl{2007SoPh..241...99M}.
\end{barticle}
\endbibitem

\bibitem[\protect\citeauthoryear{{Mari{\v{c}}i{\'c}} et~al.}{2020}]{maricic20}
\begin{barticle}
\bauthor{\bsnm{{Mari{\v{c}}i{\'c}}}, \binits{D.}},
\bauthor{\bsnm{{Vr{\v{s}}nak}}, \binits{B.}},
\bauthor{\bsnm{{Veronig}}, \binits{A.M.}},
\bauthor{\bsnm{{Dumbovi{\'c}}}, \binits{M.}},
\bauthor{\bsnm{{{\v{S}}terc}}, \binits{F.}},
\bauthor{\bsnm{{Ro{\v{s}}a}}, \binits{D.}},
\bauthor{\bsnm{{Karlica}}, \binits{M.}},
\bauthor{\bsnm{{Hr{\v{z}}ina}}, \binits{D.}},
\bauthor{\bsnm{{Rom{\v{s}}tajn}}, \binits{I.}}:
\byear{2020},
\batitle{{Sun-to-Earth Observations and Characteristics of Isolated
  Earth-Impacting Interplanetary Coronal Mass Ejections During 2008 - 2014}}.
\bjtitle{\solphys}
\bvolume{295},
\bfpage{91}.
\doiurl{https://doi.org/10.1007/s11207-020-01658-4}.
\adsurl{2020SoPh..295...91M}.
\end{barticle}
\endbibitem

\bibitem[\protect\citeauthoryear{{Mikic} and {Linker}}{1994}]{mikic94}
\begin{barticle}
\bauthor{\bsnm{{Mikic}}, \binits{Z.}},
\bauthor{\bsnm{{Linker}}, \binits{J.A.}}:
\byear{1994},
\batitle{{Disruption of Coronal Magnetic Field Arcades}}.
\bjtitle{\apj}
\bvolume{430},
\bfpage{898}.
\doiurl{https://doi.org/10.1086/174460}.
\adsurl{1994ApJ...430..898M}.
\end{barticle}
\endbibitem

\bibitem[\protect\citeauthoryear{{Miklenic}, {Veronig}, and
  {Vr{\v{s}}nak}}{2009}]{miklenic09}
\begin{barticle}
\bauthor{\bsnm{{Miklenic}}, \binits{C.H.}},
\bauthor{\bsnm{{Veronig}}, \binits{A.M.}},
\bauthor{\bsnm{{Vr{\v{s}}nak}}, \binits{B.}}:
\byear{2009},
\batitle{{Temporal comparison of nonthermal flare emission and magnetic-flux
  change rates}}.
\bjtitle{\aap}
\bvolume{499},
\bfpage{893}.
\doiurl{https://doi.org/10.1051/0004-6361/200810947}.
\adsurl{2009A&A...499..893M}.
\end{barticle}
\endbibitem

\bibitem[\protect\citeauthoryear{{Moore} et~al.}{2001}]{moore01}
\begin{barticle}
\bauthor{\bsnm{{Moore}}, \binits{R.L.}},
\bauthor{\bsnm{{Sterling}}, \binits{A.C.}},
\bauthor{\bsnm{{Hudson}}, \binits{H.S.}},
\bauthor{\bsnm{{Lemen}}, \binits{J.R.}}:
\byear{2001},
\batitle{{Onset of the Magnetic Explosion in Solar Flares and Coronal Mass
  Ejections}}.
\bjtitle{\apj}
\bvolume{552},
\bfpage{833}.
\doiurl{https://doi.org/10.1086/320559}.
\adsurl{2001ApJ...552..833M}.
\end{barticle}
\endbibitem

\bibitem[\protect\citeauthoryear{{Nieves-Chinchilla} et~al.}{2018}]{nieves18}
\begin{barticle}
\bauthor{\bsnm{{Nieves-Chinchilla}}, \binits{T.}},
\bauthor{\bsnm{{Vourlidas}}, \binits{A.}},
\bauthor{\bsnm{{Raymond}}, \binits{J.C.}},
\bauthor{\bsnm{{Linton}}, \binits{M.G.}},
\bauthor{\bsnm{{Al-haddad}}, \binits{N.}},
\bauthor{\bsnm{{Savani}}, \binits{N.P.}},
\bauthor{\bsnm{{Szabo}}, \binits{A.}},
\bauthor{\bsnm{{Hidalgo}}, \binits{M.A.}}:
\byear{2018},
\batitle{{Understanding the Internal Magnetic Field Configurations of ICMEs
  Using More than 20 Years of Wind Observations}}.
\bjtitle{\solphys}
\bvolume{293},
\bfpage{25}.
\doiurl{https://doi.org/10.1007/s11207-018-1247-z}.
\adsurl{2018SoPh..293...25N}.
\end{barticle}
\endbibitem

\bibitem[\protect\citeauthoryear{{Ouyang}, {Yang}, and {Chen}}{2015}]{ouyang15}
\begin{barticle}
\bauthor{\bsnm{{Ouyang}}, \binits{Y.}},
\bauthor{\bsnm{{Yang}}, \binits{K.}},
\bauthor{\bsnm{{Chen}}, \binits{P.F.}}:
\byear{2015},
\batitle{{Is Flux Rope a Necessary Condition for the Progenitor of Coronal Mass
  Ejections?}}
\bjtitle{\apj}
\bvolume{815},
\bfpage{72}.
\doiurl{https://doi.org/10.1088/0004-637X/815/1/72}.
\adsurl{2015ApJ...815...72O}.
\end{barticle}
\endbibitem

\bibitem[\protect\citeauthoryear{{Owens}}{2018}]{owens18}
\begin{barticle}
\bauthor{\bsnm{{Owens}}, \binits{M.J.}}:
\byear{2018},
\batitle{{Solar Wind and Heavy Ion Properties of Interplanetary Coronal Mass
  Ejections}}.
\bjtitle{\solphys}
\bvolume{293},
\bfpage{122}.
\doiurl{https://doi.org/10.1007/s11207-018-1343-0}.
\adsurl{2018SoPh..293..122O}.
\end{barticle}
\endbibitem

\bibitem[\protect\citeauthoryear{{Owocki}, {Holzer}, and
  {Hundhausen}}{1983}]{owocki83}
\begin{barticle}
\bauthor{\bsnm{{Owocki}}, \binits{S.P.}},
\bauthor{\bsnm{{Holzer}}, \binits{T.E.}},
\bauthor{\bsnm{{Hundhausen}}, \binits{A.J.}}:
\byear{1983},
\batitle{{The solar wind ionization state as a coronal temperature
  diagnostic}}.
\bjtitle{\apj}
\bvolume{275},
\bfpage{354}.
\doiurl{https://doi.org/10.1086/161538}.
\adsurl{1983ApJ...275..354O}.
\end{barticle}
\endbibitem

\bibitem[\protect\citeauthoryear{{Patsourakos}, {Vourlidas}, and
  {Stenborg}}{2013}]{patsourakos13}
\begin{barticle}
\bauthor{\bsnm{{Patsourakos}}, \binits{S.}},
\bauthor{\bsnm{{Vourlidas}}, \binits{A.}},
\bauthor{\bsnm{{Stenborg}}, \binits{G.}}:
\byear{2013},
\batitle{{Direct Evidence for a Fast Coronal Mass Ejection Driven by the Prior
  Formation and Subsequent Destabilization of a Magnetic Flux Rope}}.
\bjtitle{\apj}
\bvolume{764},
\bfpage{125}.
\doiurl{https://doi.org/10.1088/0004-637X/764/2/125}.
\adsurl{2013ApJ...764..125P}.
\end{barticle}
\endbibitem

\bibitem[\protect\citeauthoryear{{Qiu} et~al.}{2004}]{qiujiong04}
\begin{barticle}
\bauthor{\bsnm{{Qiu}}, \binits{J.}},
\bauthor{\bsnm{{Wang}}, \binits{H.}},
\bauthor{\bsnm{{Cheng}}, \binits{C.Z.}},
\bauthor{\bsnm{{Gary}}, \binits{D.E.}}:
\byear{2004},
\batitle{{Magnetic Reconnection and Mass Acceleration in Flare-Coronal Mass
  Ejection Events}}.
\bjtitle{\apj}
\bvolume{604},
\bfpage{900}.
\doiurl{https://doi.org/10.1086/382122}.
\adsurl{2004ApJ...604..900Q}.
\end{barticle}
\endbibitem

\bibitem[\protect\citeauthoryear{{Rakowski}, {Laming}, and
  {Lepri}}{2007}]{rakowski07}
\begin{barticle}
\bauthor{\bsnm{{Rakowski}}, \binits{C.E.}},
\bauthor{\bsnm{{Laming}}, \binits{J.M.}},
\bauthor{\bsnm{{Lepri}}, \binits{S.T.}}:
\byear{2007},
\batitle{{Ion Charge States in Halo Coronal Mass Ejections: What Can We Learn
  about the Explosion?}}
\bjtitle{\apj}
\bvolume{667},
\bfpage{602}.
\doiurl{https://doi.org/10.1086/520914}.
\adsurl{2007ApJ...667..602R}.
\end{barticle}
\endbibitem

\bibitem[\protect\citeauthoryear{{Ramaty} et~al.}{1995}]{ramaty95}
\begin{barticle}
\bauthor{\bsnm{{Ramaty}}, \binits{R.}},
\bauthor{\bsnm{{Mandzhavidze}}, \binits{N.}},
\bauthor{\bsnm{{Kozlovsky}}, \binits{B.}},
\bauthor{\bsnm{{Murphy}}, \binits{R.J.}}:
\byear{1995},
\batitle{{Solar Atmopheric Abundances and Energy Content in Flare Accelerated
  Ions from Gamma-Ray Spectroscopy}}.
\bjtitle{\apjl}
\bvolume{455},
\bfpage{L193}.
\doiurl{https://doi.org/10.1086/309841}.
\adsurl{1995ApJ...455L.193R}.
\end{barticle}
\endbibitem

\bibitem[\protect\citeauthoryear{{Richardson} and {Cane}}{2010}]{richardson10}
\begin{barticle}
\bauthor{\bsnm{{Richardson}}, \binits{I.G.}},
\bauthor{\bsnm{{Cane}}, \binits{H.V.}}:
\byear{2010},
\batitle{{Near-Earth Interplanetary Coronal Mass Ejections During Solar Cycle
  23 (1996 - 2009): Catalog and Summary of Properties}}.
\bjtitle{\solphys}
\bvolume{264},
\bfpage{189}.
\doiurl{https://doi.org/10.1007/s11207-010-9568-6}.
\adsurl{2010SoPh..264..189R}.
\end{barticle}
\endbibitem

\bibitem[\protect\citeauthoryear{{Rivera}, {Landi}, and
  {Lepri}}{2019}]{rivera19a}
\begin{barticle}
\bauthor{\bsnm{{Rivera}}, \binits{Y.J.}},
\bauthor{\bsnm{{Landi}}, \binits{E.}},
\bauthor{\bsnm{{Lepri}}, \binits{S.T.}}:
\byear{2019},
\batitle{{Identifying Spectral Lines to Study Coronal Mass Ejection Evolution
  in the Lower Corona}}.
\bjtitle{\apjs}
\bvolume{243},
\bfpage{34}.
\doiurl{https://doi.org/10.3847/1538-4365/ab2bfe}.
\adsurl{2019ApJS..243...34R}.
\end{barticle}
\endbibitem

\bibitem[\protect\citeauthoryear{{Rivera} et~al.}{2019}]{rivera19b}
\begin{barticle}
\bauthor{\bsnm{{Rivera}}, \binits{Y.J.}},
\bauthor{\bsnm{{Landi}}, \binits{E.}},
\bauthor{\bsnm{{Lepri}}, \binits{S.T.}},
\bauthor{\bsnm{{Gilbert}}, \binits{J.A.}}:
\byear{2019},
\batitle{{Empirical Modeling of CME Evolution Constrained to ACE/SWICS Charge
  State Distributions}}.
\bjtitle{\apj}
\bvolume{874},
\bfpage{164}.
\doiurl{https://doi.org/10.3847/1538-4357/ab0e11}.
\adsurl{2019ApJ...874..164R}.
\end{barticle}
\endbibitem

\bibitem[\protect\citeauthoryear{{Schmelz}}{1993}]{schmelz93}
\begin{barticle}
\bauthor{\bsnm{{Schmelz}}, \binits{J.T.}}:
\byear{1993},
\batitle{{Elemental Abundances of Flaring Solar Plasma: Enhanced Neon and
  Sulfur}}.
\bjtitle{\apj}
\bvolume{408},
\bfpage{373}.
\doiurl{https://doi.org/10.1086/172594}.
\adsurl{1993ApJ...408..373S}.
\end{barticle}
\endbibitem

\bibitem[\protect\citeauthoryear{{Schmelz} et~al.}{2012}]{schmelz12}
\begin{barticle}
\bauthor{\bsnm{{Schmelz}}, \binits{J.T.}},
\bauthor{\bsnm{{Reames}}, \binits{D.V.}},
\bauthor{\bsnm{{von Steiger}}, \binits{R.}},
\bauthor{\bsnm{{Basu}}, \binits{S.}}:
\byear{2012},
\batitle{{Composition of the Solar Corona, Solar Wind, and Solar Energetic
  Particles}}.
\bjtitle{\apj}
\bvolume{755},
\bfpage{33}.
\doiurl{https://doi.org/10.1088/0004-637X/755/1/33}.
\adsurl{2012ApJ...755...33S}.
\end{barticle}
\endbibitem

\bibitem[\protect\citeauthoryear{{Sharma} and {Srivastava}}{2012}]{sharma12}
\begin{barticle}
\bauthor{\bsnm{{Sharma}}, \binits{R.}},
\bauthor{\bsnm{{Srivastava}}, \binits{N.}}:
\byear{2012},
\batitle{{Presence of solar filament plasma detected in interplanetary coronal
  mass ejections by in situ spacecraft}}.
\bjtitle{Journal of Space Weather and Space Climate}
\bvolume{2},
\bfpage{A10}.
\doiurl{https://doi.org/10.1051/swsc/2012010}.
\adsurl{2012JSWSC...2A..10S}.
\end{barticle}
\endbibitem

\bibitem[\protect\citeauthoryear{{Shearer} et~al.}{2014}]{shearer14}
\begin{barticle}
\bauthor{\bsnm{{Shearer}}, \binits{P.}},
\bauthor{\bsnm{{von Steiger}}, \binits{R.}},
\bauthor{\bsnm{{Raines}}, \binits{J.M.}},
\bauthor{\bsnm{{Lepri}}, \binits{S.T.}},
\bauthor{\bsnm{{Thomas}}, \binits{J.W.}},
\bauthor{\bsnm{{Gilbert}}, \binits{J.A.}},
\bauthor{\bsnm{{Landi}}, \binits{E.}},
\bauthor{\bsnm{{Zurbuchen}}, \binits{T.H.}}:
\byear{2014},
\batitle{{The Solar Wind Neon Abundance Observed with ACE/SWICS and
  Ulysses/SWICS}}.
\bjtitle{\apj}
\bvolume{789},
\bfpage{60}.
\doiurl{https://doi.org/10.1088/0004-637X/789/1/60}.
\adsurl{2014ApJ...789...60S}.
\end{barticle}
\endbibitem

\bibitem[\protect\citeauthoryear{{Shemi}}{1991}]{shemi91}
\begin{barticle}
\bauthor{\bsnm{{Shemi}}, \binits{A.}}:
\byear{1991},
\batitle{{The high Ne/C and Ne/O abundance ratios in the footpoints of solar
  flares}}.
\bjtitle{\mnras}
\bvolume{251},
\bfpage{221}.
\doiurl{https://doi.org/10.1093/mnras/251.2.221}.
\adsurl{1991MNRAS.251..221S}.
\end{barticle}
\endbibitem

\bibitem[\protect\citeauthoryear{{Shen} et~al.}{2017}]{shenchenglong17}
\begin{barticle}
\bauthor{\bsnm{{Shen}}, \binits{C.}},
\bauthor{\bsnm{{Chi}}, \binits{Y.}},
\bauthor{\bsnm{{Wang}}, \binits{Y.}},
\bauthor{\bsnm{{Xu}}, \binits{M.}},
\bauthor{\bsnm{{Wang}}, \binits{S.}}:
\byear{2017},
\batitle{{Statistical comparison of the ICME's geoeffectiveness of different
  types and different solar phases from 1995 to 2014}}.
\bjtitle{Journal of Geophysical Research (Space Physics)}
\bvolume{122},
\bfpage{5931}.
\doiurl{https://doi.org/10.1002/2016JA023768}.
\adsurl{2017JGRA..122.5931S}.
\end{barticle}
\endbibitem

\bibitem[\protect\citeauthoryear{{Shi} et~al.}{2019}]{shimijie19}
\begin{barticle}
\bauthor{\bsnm{{Shi}}, \binits{M.}},
\bauthor{\bsnm{{Li}}, \binits{B.}},
\bauthor{\bsnm{{Van Doorsselaere}}, \binits{T.}},
\bauthor{\bsnm{{Chen}}, \binits{S.-X.}},
\bauthor{\bsnm{{Huang}}, \binits{Z.}}:
\byear{2019},
\batitle{{Non-equilibrium Ionization Effects on Extreme-ultraviolet Emissions
  Modulated by Standing Sausage Modes in Coronal Loops}}.
\bjtitle{\apj}
\bvolume{870},
\bfpage{99}.
\doiurl{https://doi.org/10.3847/1538-4357/aaf393}.
\adsurl{2019ApJ...870...99S}.
\end{barticle}
\endbibitem

\bibitem[\protect\citeauthoryear{{Song} and {Yao}}{2020}]{song20b}
\begin{barticle}
\bauthor{\bsnm{{Song}}, \binits{H.}},
\bauthor{\bsnm{{Yao}}, \binits{S.}}:
\byear{2020},
\batitle{{Characteristics and applications of interplanetary coronal mass
  ejection composition}}.
\bjtitle{Sci China Tech Sci}
\bvolume{63},
\bfpage{2171}.
\doiurl{https://doi.org/10.1007/s11431-020-1680-y}.
\adsurl{2020arXiv200611473S}.
\end{barticle}
\endbibitem

\bibitem[\protect\citeauthoryear{{Song} et~al.}{2013}]{song13}
\begin{barticle}
\bauthor{\bsnm{{Song}}, \binits{H.Q.}},
\bauthor{\bsnm{{Chen}}, \binits{Y.}},
\bauthor{\bsnm{{Ye}}, \binits{D.D.}},
\bauthor{\bsnm{{Han}}, \binits{G.Q.}},
\bauthor{\bsnm{{Du}}, \binits{G.H.}},
\bauthor{\bsnm{{Li}}, \binits{G.}},
\bauthor{\bsnm{{Zhang}}, \binits{J.}},
\bauthor{\bsnm{{Hu}}, \binits{Q.}}:
\byear{2013},
\batitle{{A Study of Fast Flareless Coronal Mass Ejections}}.
\bjtitle{\apj}
\bvolume{773},
\bfpage{129}.
\doiurl{https://doi.org/10.1088/0004-637X/773/2/129}.
\adsurl{2013ApJ...773..129S}.
\end{barticle}
\endbibitem

\bibitem[\protect\citeauthoryear{{Song} et~al.}{2014}]{song14a}
\begin{barticle}
\bauthor{\bsnm{{Song}}, \binits{H.Q.}},
\bauthor{\bsnm{{Zhang}}, \binits{J.}},
\bauthor{\bsnm{{Chen}}, \binits{Y.}},
\bauthor{\bsnm{{Cheng}}, \binits{X.}}:
\byear{2014},
\batitle{{Direct Observations of Magnetic Flux Rope Formation during a Solar
  Coronal Mass Ejection}}.
\bjtitle{\apjl}
\bvolume{792},
\bfpage{L40}.
\doiurl{https://doi.org/10.1088/2041-8205/792/2/L40}.
\adsurl{2014ApJ...792L..40S}.
\end{barticle}
\endbibitem

\bibitem[\protect\citeauthoryear{{Song} et~al.}{2015a}]{song15a}
\begin{barticle}
\bauthor{\bsnm{{Song}}, \binits{H.Q.}},
\bauthor{\bsnm{{Chen}}, \binits{Y.}},
\bauthor{\bsnm{{Zhang}}, \binits{J.}},
\bauthor{\bsnm{{Cheng}}, \binits{X.}},
\bauthor{\bsnm{{Fu}}, \binits{H.}},
\bauthor{\bsnm{{LI}}, \binits{G.}}:
\byear{2015}a,
\batitle{{Acceleration Phases of a Solar Filament During Its Eruption}}.
\bjtitle{\apjl}
\bvolume{804},
\bfpage{L38}.
\doiurl{https://doi.org/10.1088/2041-8205/804/2/L38}.
\adsurl{2015ApJ...804L..38S}.
\end{barticle}
\endbibitem

\bibitem[\protect\citeauthoryear{{Song} et~al.}{2015b}]{song15b}
\begin{barticle}
\bauthor{\bsnm{{Song}}, \binits{H.Q.}},
\bauthor{\bsnm{{Chen}}, \binits{Y.}},
\bauthor{\bsnm{{Zhang}}, \binits{J.}},
\bauthor{\bsnm{{Cheng}}, \binits{X.}},
\bauthor{\bsnm{{Wang}}, \binits{B.}},
\bauthor{\bsnm{{Hu}}, \binits{Q.}},
\bauthor{\bsnm{{Li}}, \binits{G.}},
\bauthor{\bsnm{{Wang}}, \binits{Y.M.}}:
\byear{2015}b,
\batitle{{Evidence of the Solar EUV Hot Channel as a Magnetic Flux Rope from
  Remote-sensing and In Situ Observations}}.
\bjtitle{\apjl}
\bvolume{808},
\bfpage{L15}.
\doiurl{https://doi.org/10.1088/2041-8205/808/1/L15}.
\adsurl{2015ApJ...808L..15S}.
\end{barticle}
\endbibitem

\bibitem[\protect\citeauthoryear{{Song} et~al.}{2015c}]{song15c}
\begin{barticle}
\bauthor{\bsnm{{Song}}, \binits{H.Q.}},
\bauthor{\bsnm{{Zhang}}, \binits{J.}},
\bauthor{\bsnm{{Chen}}, \binits{Y.}},
\bauthor{\bsnm{{Cheng}}, \binits{X.}},
\bauthor{\bsnm{{Li}}, \binits{G.}},
\bauthor{\bsnm{{Wang}}, \binits{Y.M.}}:
\byear{2015}c,
\batitle{{First Taste of Hot Channel in Interplanetary Space}}.
\bjtitle{\apj}
\bvolume{803},
\bfpage{96}.
\doiurl{https://doi.org/10.1088/0004-637X/803/2/96}.
\adsurl{2015ApJ...803...96S}.
\end{barticle}
\endbibitem

\bibitem[\protect\citeauthoryear{{Song} et~al.}{2016}]{song16}
\begin{barticle}
\bauthor{\bsnm{{Song}}, \binits{H.Q.}},
\bauthor{\bsnm{{Zhong}}, \binits{Z.}},
\bauthor{\bsnm{{Chen}}, \binits{Y.}},
\bauthor{\bsnm{{Zhang}}, \binits{J.}},
\bauthor{\bsnm{{Cheng}}, \binits{X.}},
\bauthor{\bsnm{{Zhao}}, \binits{L.}},
\bauthor{\bsnm{{Hu}}, \binits{Q.}},
\bauthor{\bsnm{{Li}}, \binits{G.}}:
\byear{2016},
\batitle{{A Statistical Study of the Average Iron Charge State Distributions
  inside Magnetic Clouds for Solar Cycle 23}}.
\bjtitle{\apjs}
\bvolume{224},
\bfpage{27}.
\doiurl{https://doi.org/10.3847/0067-0049/224/2/27}.
\adsurl{2016ApJS..224...27S}.
\end{barticle}
\endbibitem

\bibitem[\protect\citeauthoryear{{Song} et~al.}{2017a}]{song17a}
\begin{barticle}
\bauthor{\bsnm{{Song}}, \binits{H.Q.}},
\bauthor{\bsnm{{Chen}}, \binits{Y.}},
\bauthor{\bsnm{{Li}}, \binits{B.}},
\bauthor{\bsnm{{Li}}, \binits{L.P.}},
\bauthor{\bsnm{{Zhao}}, \binits{L.}},
\bauthor{\bsnm{{He}}, \binits{J.S.}},
\bauthor{\bsnm{{Duan}}, \binits{D.}},
\bauthor{\bsnm{{Cheng}}, \binits{X.}},
\bauthor{\bsnm{{Zhang}}, \binits{J.}}:
\byear{2017}a,
\batitle{{The Origin of Solar Filament Plasma Inferred from In Situ
  Observations of Elemental Abundances}}.
\bjtitle{\apjl}
\bvolume{836},
\bfpage{L11}.
\doiurl{https://doi.org/10.3847/2041-8213/aa5d54}.
\adsurl{2017ApJ...836L..11S}.
\end{barticle}
\endbibitem

\bibitem[\protect\citeauthoryear{{Song} et~al.}{2017b}]{song17b}
\begin{barticle}
\bauthor{\bsnm{{Song}}, \binits{H.Q.}},
\bauthor{\bsnm{{Cheng}}, \binits{X.}},
\bauthor{\bsnm{{Chen}}, \binits{Y.}},
\bauthor{\bsnm{{Zhang}}, \binits{J.}},
\bauthor{\bsnm{{Wang}}, \binits{B.}},
\bauthor{\bsnm{{Li}}, \binits{L.P.}},
\bauthor{\bsnm{{Li}}, \binits{B.}},
\bauthor{\bsnm{{Hu}}, \binits{Q.}},
\bauthor{\bsnm{{Li}}, \binits{G.}}:
\byear{2017}b,
\batitle{{The Three-part Structure of a Filament-unrelated Solar Coronal Mass
  Ejection}}.
\bjtitle{\apj}
\bvolume{848},
\bfpage{21}.
\doiurl{https://doi.org/10.3847/1538-4357/aa8d1a}.
\adsurl{2017ApJ...848...21S}.
\end{barticle}
\endbibitem

\bibitem[\protect\citeauthoryear{{Song} et~al.}{2018}]{song18a}
\begin{barticle}
\bauthor{\bsnm{{Song}}, \binits{H.Q.}},
\bauthor{\bsnm{{Chen}}, \binits{Y.}},
\bauthor{\bsnm{{Qiu}}, \binits{J.}},
\bauthor{\bsnm{{Chen}}, \binits{C.X.}},
\bauthor{\bsnm{{Zhang}}, \binits{J.}},
\bauthor{\bsnm{{Cheng}}, \binits{X.}},
\bauthor{\bsnm{{Shen}}, \binits{Y.D.}},
\bauthor{\bsnm{{Zheng}}, \binits{R.S.}}:
\byear{2018},
\batitle{{The Acceleration Process of a Solar Quiescent Filament in the Inner
  Corona}}.
\bjtitle{\apjl}
\bvolume{857},
\bfpage{L21}.
\doiurl{https://doi.org/10.3847/2041-8213/aabcc3}.
\adsurl{2018ApJ...857L..21S}.
\end{barticle}
\endbibitem

\bibitem[\protect\citeauthoryear{{Song} et~al.}{2019a}]{song19a}
\begin{barticle}
\bauthor{\bsnm{{Song}}, \binits{H.Q.}},
\bauthor{\bsnm{{Zhang}}, \binits{J.}},
\bauthor{\bsnm{{Cheng}}, \binits{X.}},
\bauthor{\bsnm{{Li}}, \binits{L.P.}},
\bauthor{\bsnm{{Tang}}, \binits{Y.Z.}},
\bauthor{\bsnm{{Wang}}, \binits{B.}},
\bauthor{\bsnm{{Zheng}}, \binits{R.S.}},
\bauthor{\bsnm{{Chen}}, \binits{Y.}}:
\byear{2019}a,
\batitle{{On the Nature of the Bright Core of Solar Coronal Mass Ejections}}.
\bjtitle{\apj}
\bvolume{883},
\bfpage{43}.
\doiurl{https://doi.org/10.3847/1538-4357/ab304c}.
\adsurl{2019ApJ...883...43S}.
\end{barticle}
\endbibitem

\bibitem[\protect\citeauthoryear{{Song} et~al.}{2019b}]{song19b}
\begin{barticle}
\bauthor{\bsnm{{Song}}, \binits{H.Q.}},
\bauthor{\bsnm{{Zhang}}, \binits{J.}},
\bauthor{\bsnm{{Li}}, \binits{L.P.}},
\bauthor{\bsnm{{Liu}}, \binits{Y.D.}},
\bauthor{\bsnm{{Zhu}}, \binits{B.}},
\bauthor{\bsnm{{Wang}}, \binits{B.}},
\bauthor{\bsnm{{Zheng}}, \binits{R.S.}},
\bauthor{\bsnm{{Chen}}, \binits{Y.}}:
\byear{2019}b,
\batitle{{The Structure of Solar Coronal Mass Ejections in the
  Extreme-ultraviolet Passbands}}.
\bjtitle{\apj}
\bvolume{887},
\bfpage{124}.
\doiurl{https://doi.org/10.3847/1538-4357/ab50b6}.
\adsurl{2019ApJ...887..124S}.
\end{barticle}
\endbibitem

\bibitem[\protect\citeauthoryear{{Song} et~al.}{2020}]{song20a}
\begin{barticle}
\bauthor{\bsnm{{Song}}, \binits{H.Q.}},
\bauthor{\bsnm{{Zhang}}, \binits{J.}},
\bauthor{\bsnm{{Cheng}}, \binits{X.}},
\bauthor{\bsnm{{Li}}, \binits{G.}},
\bauthor{\bsnm{{Hu}}, \binits{Q.}},
\bauthor{\bsnm{{Li}}, \binits{L.P.}},
\bauthor{\bsnm{{Chen}}, \binits{S.J.}},
\bauthor{\bsnm{{Zheng}}, \binits{R.S.}},
\bauthor{\bsnm{{Chen}}, \binits{Y.}}:
\byear{2020},
\batitle{{Do All Interplanetary Coronal Mass Ejections Have a Magnetic Flux
  Rope Structure Near 1 au?}}
\bjtitle{\apjl}
\bvolume{901},
\bfpage{L21}.
\doiurl{https://doi.org/10.3847/2041-8213/abb6ec}.
\adsurl{2020ApJ...901L..21S}.
\end{barticle}
\endbibitem

\bibitem[\protect\citeauthoryear{{Vourlidas} et~al.}{2013}]{vourlidas13}
\begin{barticle}
\bauthor{\bsnm{{Vourlidas}}, \binits{A.}},
\bauthor{\bsnm{{Lynch}}, \binits{B.J.}},
\bauthor{\bsnm{{Howard}}, \binits{R.A.}},
\bauthor{\bsnm{{Li}}, \binits{Y.}}:
\byear{2013},
\batitle{{How Many CMEs Have Flux Ropes? Deciphering the Signatures of Shocks,
  Flux Ropes, and Prominences in Coronagraph Observations of CMEs}}.
\bjtitle{\solphys}
\bvolume{284},
\bfpage{179}.
\doiurl{https://doi.org/10.1007/s11207-012-0084-8}.
\adsurl{2013SoPh..284..179V}.
\end{barticle}
\endbibitem

\bibitem[\protect\citeauthoryear{{Wang}, {Feng}, and
  {Zhao}}{2018}]{wangjiemin18}
\begin{barticle}
\bauthor{\bsnm{{Wang}}, \binits{J.}},
\bauthor{\bsnm{{Feng}}, \binits{H.}},
\bauthor{\bsnm{{Zhao}}, \binits{G.}}:
\byear{2018},
\batitle{{Cold prominence materials detected within magnetic clouds during
  1998-2007}}.
\bjtitle{\aap}
\bvolume{616},
\bfpage{A41}.
\doiurl{https://doi.org/10.1051/0004-6361/201731807}.
\adsurl{2018A&A...616A..41W}.
\end{barticle}
\endbibitem

\bibitem[\protect\citeauthoryear{{Wang} et~al.}{2017}]{wangwensi17}
\begin{barticle}
\bauthor{\bsnm{{Wang}}, \binits{W.}},
\bauthor{\bsnm{{Liu}}, \binits{R.}},
\bauthor{\bsnm{{Wang}}, \binits{Y.}},
\bauthor{\bsnm{{Hu}}, \binits{Q.}},
\bauthor{\bsnm{{Shen}}, \binits{C.}},
\bauthor{\bsnm{{Jiang}}, \binits{C.}},
\bauthor{\bsnm{{Zhu}}, \binits{C.}}:
\byear{2017},
\batitle{{Buildup of a highly twisted magnetic flux rope during a solar
  eruption}}.
\bjtitle{Nature Communications}
\bvolume{8},
\bfpage{1330}.
\doiurl{https://doi.org/10.1038/s41467-017-01207-x}.
\adsurl{2017NatCo...8.1330W}.
\end{barticle}
\endbibitem

\bibitem[\protect\citeauthoryear{{Webb} and {Howard}}{2012}]{webb12}
\begin{barticle}
\bauthor{\bsnm{{Webb}}, \binits{D.F.}},
\bauthor{\bsnm{{Howard}}, \binits{T.A.}}:
\byear{2012},
\batitle{{Coronal Mass Ejections: Observations}}.
\bjtitle{Living Reviews in Solar Physics}
\bvolume{9},
\bfpage{3}.
\doiurl{https://doi.org/10.12942/lrsp-2012-3}.
\adsurl{2012LRSP....9....3W}.
\end{barticle}
\endbibitem

\bibitem[\protect\citeauthoryear{{Xie}, {Gopalswamy}, and {St.
  Cyr}}{2013}]{xie13}
\begin{barticle}
\bauthor{\bsnm{{Xie}}, \binits{H.}},
\bauthor{\bsnm{{Gopalswamy}}, \binits{N.}},
\bauthor{\bsnm{{St. Cyr}}, \binits{O.C.}}:
\byear{2013},
\batitle{{Near-Sun Flux-Rope Structure of CMEs}}.
\bjtitle{\solphys}
\bvolume{284},
\bfpage{47}.
\doiurl{https://doi.org/10.1007/s11207-012-0209-0}.
\adsurl{2013SoPh..284...47X}.
\end{barticle}
\endbibitem

\bibitem[\protect\citeauthoryear{{Yeh} and {Lindau}}{1985}]{yeh85}
\begin{barticle}
\bauthor{\bsnm{{Yeh}}, \binits{J.J.}},
\bauthor{\bsnm{{Lindau}}, \binits{I.}}:
\byear{1985},
\batitle{{Atomic Subshell Photoionization Cross Sections and Asymmetry
  Parameters: 1 <= Z <= 103}}.
\bjtitle{Atomic Data and Nuclear Data Tables}
\bvolume{32},
\bfpage{1}.
\doiurl{https://doi.org/10.1016/0092-640X(85)90016-6}.
\adsurl{1985ADNDT..32....1Y}.
\end{barticle}
\endbibitem

\bibitem[\protect\citeauthoryear{{Young}}{2005}]{young05}
\begin{barticle}
\bauthor{\bsnm{{Young}}, \binits{P.R.}}:
\byear{2005},
\batitle{{The element abundance FIP effect in the quiet Sun}}.
\bjtitle{\aap}
\bvolume{439},
\bfpage{361}.
\doiurl{https://doi.org/10.1051/0004-6361:20052963}.
\adsurl{2005A&A...439..361Y}.
\end{barticle}
\endbibitem

\bibitem[\protect\citeauthoryear{{Zhang}, {Hess}, and
  {Poomvises}}{2013}]{zhangjie13}
\begin{barticle}
\bauthor{\bsnm{{Zhang}}, \binits{J.}},
\bauthor{\bsnm{{Hess}}, \binits{P.}},
\bauthor{\bsnm{{Poomvises}}, \binits{W.}}:
\byear{2013},
\batitle{{A Comparative Study of Coronal Mass Ejections with and Without
  Magnetic Cloud Structure near the Earth: Are All Interplanetary CMEs Flux
  Ropes?}}
\bjtitle{\solphys}
\bvolume{284},
\bfpage{89}.
\doiurl{https://doi.org/10.1007/s11207-013-0242-7}.
\adsurl{2013SoPh..284...89Z}.
\end{barticle}
\endbibitem

\bibitem[\protect\citeauthoryear{{Zhang} et~al.}{2001}]{zhangjie01}
\begin{barticle}
\bauthor{\bsnm{{Zhang}}, \binits{J.}},
\bauthor{\bsnm{{Dere}}, \binits{K.P.}},
\bauthor{\bsnm{{Howard}}, \binits{R.A.}},
\bauthor{\bsnm{{Kundu}}, \binits{M.R.}},
\bauthor{\bsnm{{White}}, \binits{S.M.}}:
\byear{2001},
\batitle{{On the Temporal Relationship between Coronal Mass Ejections and
  Flares}}.
\bjtitle{\apj}
\bvolume{559},
\bfpage{452}.
\doiurl{https://doi.org/10.1086/322405}.
\adsurl{2001ApJ...559..452Z}.
\end{barticle}
\endbibitem

\bibitem[\protect\citeauthoryear{{Zhang} et~al.}{2007}]{zhangjie07}
\begin{barticle}
\bauthor{\bsnm{{Zhang}}, \binits{J.}},
\bauthor{\bsnm{{Richardson}}, \binits{I.G.}},
\bauthor{\bsnm{{Webb}}, \binits{D.F.}},
\bauthor{\bsnm{{Gopalswamy}}, \binits{N.}},
\bauthor{\bsnm{{Huttunen}}, \binits{E.}},
\bauthor{\bsnm{{Kasper}}, \binits{J.C.}},
\bauthor{\bsnm{{Nitta}}, \binits{N.V.}},
\bauthor{\bsnm{{Poomvises}}, \binits{W.}},
\bauthor{\bsnm{{Thompson}}, \binits{B.J.}},
\bauthor{\bsnm{{Wu}}, \binits{C.-C.}},
\bauthor{\bsnm{{Yashiro}}, \binits{S.}},
\bauthor{\bsnm{{Zhukov}}, \binits{A.N.}}:
\byear{2007},
\batitle{{Solar and interplanetary sources of major geomagnetic storms (Dst <=
  -100 nT) during 1996-2005}}.
\bjtitle{Journal of Geophysical Research (Space Physics)}
\bvolume{112},
\bfpage{A10102}.
\doiurl{https://doi.org/10.1029/2007JA012321}.
\adsurl{2007JGRA..11210102Z}.
\end{barticle}
\endbibitem

\bibitem[\protect\citeauthoryear{{Zhao} et~al.}{2014}]{zhaoliang14}
\begin{barticle}
\bauthor{\bsnm{{Zhao}}, \binits{L.}},
\bauthor{\bsnm{{Landi}}, \binits{E.}},
\bauthor{\bsnm{{Zurbuchen}}, \binits{T.H.}},
\bauthor{\bsnm{{Fisk}}, \binits{L.A.}},
\bauthor{\bsnm{{Lepri}}, \binits{S.T.}}:
\byear{2014},
\batitle{{The Evolution of 1 AU Equatorial Solar Wind and its Association with
  the Morphology of the Heliospheric Current Sheet from Solar Cycles 23 to
  24}}.
\bjtitle{\apj}
\bvolume{793},
\bfpage{44}.
\doiurl{https://doi.org/10.1088/0004-637X/793/1/44}.
\adsurl{2014ApJ...793...44Z}.
\end{barticle}
\endbibitem

\bibitem[\protect\citeauthoryear{{Zhao} et~al.}{2017a}]{zhaoliang17a}
\begin{barticle}
\bauthor{\bsnm{{Zhao}}, \binits{L.}},
\bauthor{\bsnm{{Landi}}, \binits{E.}},
\bauthor{\bsnm{{Lepri}}, \binits{S.T.}},
\bauthor{\bsnm{{Kocher}}, \binits{M.}},
\bauthor{\bsnm{{Zurbuchen}}, \binits{T.H.}},
\bauthor{\bsnm{{Fisk}}, \binits{L.A.}},
\bauthor{\bsnm{{Raines}}, \binits{J.M.}}:
\byear{2017}a,
\batitle{{An Anomalous Composition in Slow Solar Wind as a Signature of
  Magnetic Reconnection in its Source Region}}.
\bjtitle{\apjs}
\bvolume{228},
\bfpage{4}.
\doiurl{https://doi.org/10.3847/1538-4365/228/1/4}.
\adsurl{2017ApJS..228....4Z}.
\end{barticle}
\endbibitem

\bibitem[\protect\citeauthoryear{{Zhao} et~al.}{2017b}]{zhaoliang17}
\begin{barticle}
\bauthor{\bsnm{{Zhao}}, \binits{L.}},
\bauthor{\bsnm{{Landi}}, \binits{E.}},
\bauthor{\bsnm{{Lepri}}, \binits{S.T.}},
\bauthor{\bsnm{{Gilbert}}, \binits{J.A.}},
\bauthor{\bsnm{{Zurbuchen}}, \binits{T.H.}},
\bauthor{\bsnm{{Fisk}}, \binits{L.A.}},
\bauthor{\bsnm{{Raines}}, \binits{J.M.}}:
\byear{2017}b,
\batitle{{On the Relation between the In Situ Properties and the Coronal
  Sources of the Solar Wind}}.
\bjtitle{\apj}
\bvolume{846},
\bfpage{135}.
\doiurl{https://doi.org/10.3847/1538-4357/aa850c}.
\adsurl{2017ApJ...846..135Z}.
\end{barticle}
\endbibitem

\bibitem[\protect\citeauthoryear{{Zhu} et~al.}{2020}]{zhuchunming20}
\begin{barticle}
\bauthor{\bsnm{{Zhu}}, \binits{C.}},
\bauthor{\bsnm{{Qiu}}, \binits{J.}},
\bauthor{\bsnm{{Liewer}}, \binits{P.}},
\bauthor{\bsnm{{Vourlidas}}, \binits{A.}},
\bauthor{\bsnm{{Spiegel}}, \binits{M.}},
\bauthor{\bsnm{{Hu}}, \binits{Q.}}:
\byear{2020},
\batitle{{How Does Magnetic Reconnection Drive the Early-stage Evolution of
  Coronal Mass Ejections?}}
\bjtitle{\apj}
\bvolume{893},
\bfpage{141}.
\doiurl{https://doi.org/10.3847/1538-4357/ab838a}.
\adsurl{2020ApJ...893..141Z}.
\end{barticle}
\endbibitem

\bibitem[\protect\citeauthoryear{{Zurbuchen} et~al.}{2002}]{zurbuchen02}
\begin{barticle}
\bauthor{\bsnm{{Zurbuchen}}, \binits{T.H.}},
\bauthor{\bsnm{{Fisk}}, \binits{L.A.}},
\bauthor{\bsnm{{Gloeckler}}, \binits{G.}},
\bauthor{\bsnm{{von Steiger}}, \binits{R.}}:
\byear{2002},
\batitle{{The solar wind composition throughout the solar cycle: A continuum of
  dynamic states}}.
\bjtitle{\grl}
\bvolume{29},
\bfpage{1352}.
\doiurl{https://doi.org/10.1029/2001GL013946}.
\adsurl{2002GeoRL..29.1352Z}.
\end{barticle}
\endbibitem

\bibitem[\protect\citeauthoryear{{Zurbuchen} et~al.}{2016}]{zurbuchen16}
\begin{barticle}
\bauthor{\bsnm{{Zurbuchen}}, \binits{T.H.}},
\bauthor{\bsnm{{Weberg}}, \binits{M.}},
\bauthor{\bsnm{{von Steiger}}, \binits{R.}},
\bauthor{\bsnm{{Mewaldt}}, \binits{R.A.}},
\bauthor{\bsnm{{Lepri}}, \binits{S.T.}},
\bauthor{\bsnm{{Antiochos}}, \binits{S.K.}}:
\byear{2016},
\batitle{{Composition of Coronal Mass Ejections}}.
\bjtitle{\apj}
\bvolume{826},
\bfpage{10}.
\doiurl{https://doi.org/10.3847/0004-637X/826/1/10}.
\adsurl{2016ApJ...826...10Z}.
\end{barticle}
\endbibitem

\end{thebibliography}

\begin{figure*}
\centering
\includegraphics[width=10cm]{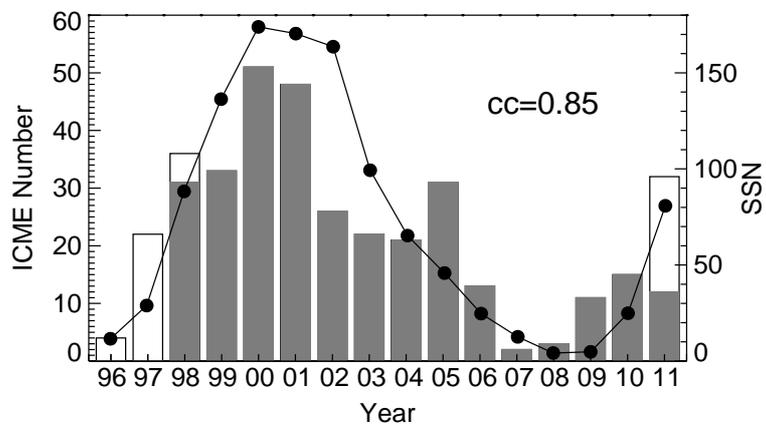} \caption{Yearly numbers of ICMEs at L1 point and sunspots from 1996 to 2011. The histograms depict the ICME numbers with the gray regions indicating the numbers during SWICS optimal science configuration. The black line connected by the filled circles displays the SSNs. The correlation coefficient (cc) between the numbers of ICMEs and sunspots is also presented. \label{Figure 1}}
\end{figure*}

\begin{figure*}
\centering
\includegraphics[width=6cm]{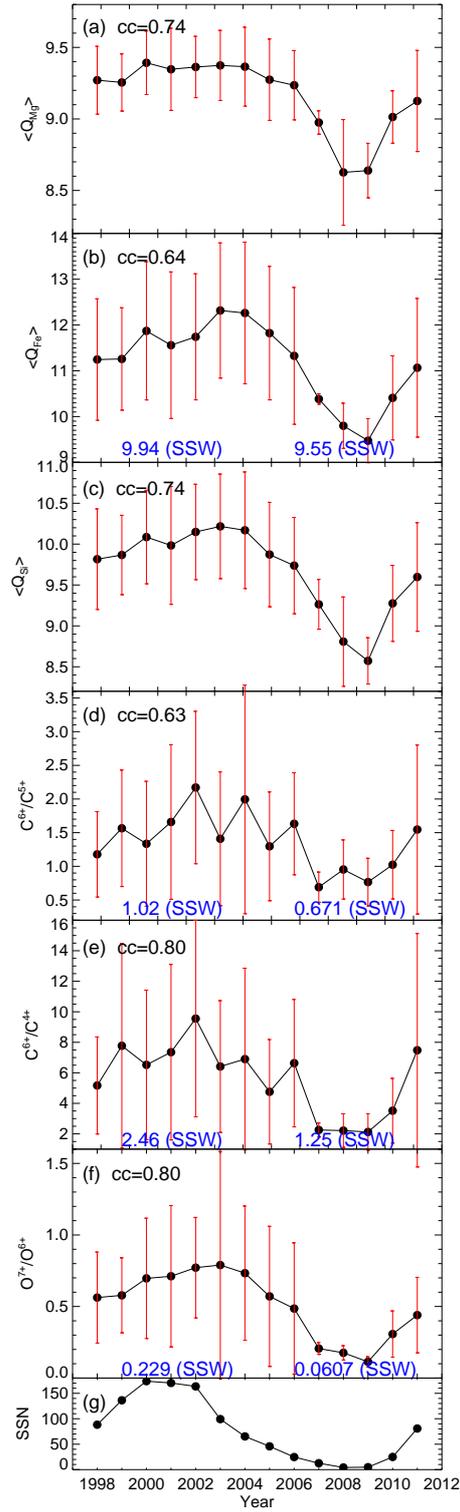} \caption{The solar cycle dependence of charge states within ICMEs from 1998 to 2011. The panels show the variations of (a) $<$Q$_{Mg}$$>$, (b) $<$Q$_{Fe}$$>$, (c) $<$Q$_{Si}$$>$, (d) C$^{6+}$/C$^{5+}$, (e) C$^{6+}$/C$^{4+}$, (f) O$^{7+}$/O$^{6+}$, and (g) yearly average SSNs. The vertical red lines show the standard deviations. The correlation coefficients (cc) between the charge states and SSNs are presented in Panels (a)--(f). The blue numbers refer to the corresponding means in the slow solar wind (SSW) during solar maximum (left) and minimum (right) \citep{lepri13}. \label{Figure 2}}
\end{figure*}

\begin{figure*}
\centering
\includegraphics[width=6cm]{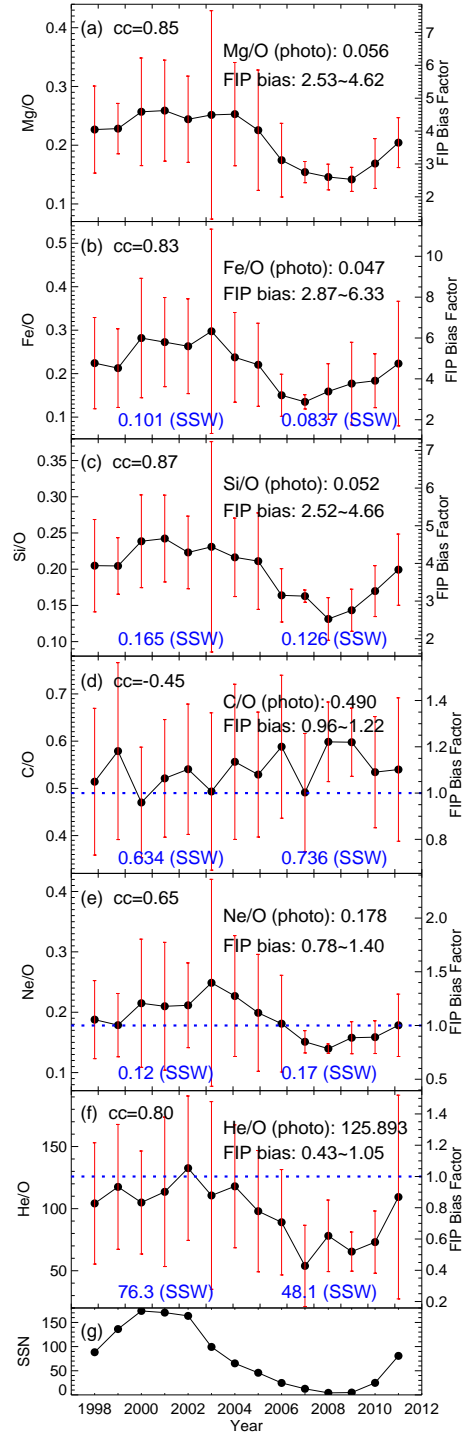} \caption{The solar cycle dependence of elemental abundances within ICMEs from 1998 to 2011. The panels show the variations of (a) Mg/O, (b) Fe/O, (c) Si/O, (d) C/O, (e) Ne/O, (f) He/O, and (g) yearly average SSNs. The vertical red lines show the standard deviations. The correlation coefficients (cc) between the abundances and SSNs are presented in Panels (a)--(f). The horizontal blue dotted lines in Panels (d)--(f) denote the corresponding values in the photosphere. The blue numbers refer to the corresponding values in the slow solar wind (SSW) during solar maximum (left) and minimum (right) \citep{lepri13,shearer14}. \label{Figure 3}}
\end{figure*}

\begin{figure*}
\centering
\includegraphics[width=12cm]{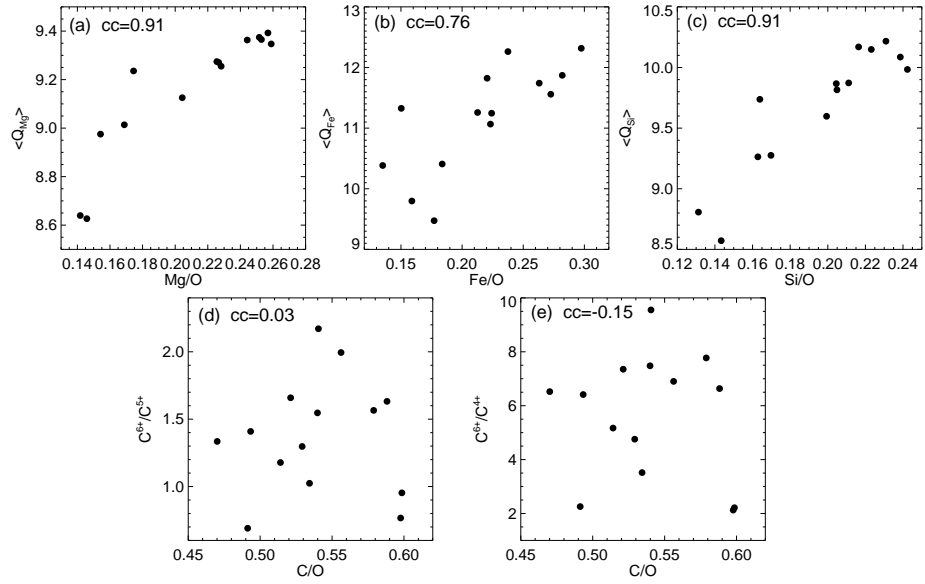} \caption{The scatter plots of yearly average values of charge states and relative elemental abundances within ICMEs. The correlation coefficients (cc) between them are presented in each panel. \label{Figure 4}}
\end{figure*}

\end{article}
\end{document}